\documentclass[aps,nofootinbib,superscriptaddress, showpacs,preprintnumbers,  twocolumn, nofootinbibt]{revtex4-2}
\usepackage{eurosym}
\usepackage{amsmath}
\usepackage{bm}
\usepackage{amsfonts}
\usepackage{amssymb}
\usepackage{float}
\usepackage{graphicx}
\usepackage{subfig}
\usepackage{caption}
\usepackage{comment}
\usepackage{multirow}
\usepackage[export]{adjustbox}
\graphicspath{ {figures} }
\usepackage{xcolor}
\usepackage{hyperref}
\usepackage{booktabs}
\usepackage{tikz}
\usetikzlibrary{arrows.meta, positioning}

\renewcommand{\notag}{\nn}

\def\be{\begin{equation}}
	\def\ee{\end{equation}}
\def\bea{\begin{eqnarray}}
	\def\eea{\end{eqnarray}}

\def\nn{\nonumber \\}

\begin{document}
	
	\title{Big Bang Nucleosynthesis constraints on the cosmological evolution in a Universe with a Weylian Boundary}
	\author{Teodora M. Matei}
	\email{teodora.maria.matei@stud.ubbcluj.ro}
	\affiliation{Department of Physics, Babe\c s-Bolyai University, 1 Kog\u alniceanu Street,
		Cluj-Napoca 400084, Romania,}
	\affiliation{Astronomical Observatory, 19 Cire\c silor Street,
			Cluj-Napoca 400487, Romania}
	\author{Cristian A. Croitoru}
	\email{croitoru.lu.cristian@student.utcluj.ro}
	\affiliation{Faculty of Computer Science and Automation, Technical University of Cluj-Napoca, G. Baritiu Street 26-28, Cluj-Napoca 400027, Romania}
	\author{Tiberiu Harko}
	\email{tiberiu.harko@aira.astro.ro}
	\affiliation{Department of Physics, Babe\c s-Bolyai University, 1 Kog\u alniceanu Street,
		Cluj-Napoca 400084, Romania,}
	\affiliation{Astronomical Observatory, 19 Cire\c silor Street,
		Cluj-Napoca 400487, Romania}

	\begin{abstract}
	We investigate the effects that arise from the inclusion of boundary terms in the Einstein gravitational field equations  in the Big Bang Nucleosynthesis (BBN) framework. In particular, we consider the possibility that the boundary of the Universe is described by a Weyl type geometry. With the help of the generalized Friedmann equations for a Universe with a Weylian boundary, obtained for a Friedmann-Lemaitre-Robertson-Walker FLRW metric, three distinct cosmological models can be constructed.  The cosmological evolution is determined by a dissipative scalar field, and by the Weyl vector coming from the boundary. Several cosmological scenarios are obtained via the appropriate splitting of the generalized energy conservation equation. In the present work we obtain relevant constraints on these models by using  the BBN data.
In particular, the effects on the BBN that arise in the post warm-inflationary era will be examined by theoretically evaluating the measured abundances of relic nuclei (Hydrogen, Deuterium, Helium-3, Helium-4, and Lithium-7).  
We consider firstly the primordial mass fraction estimates, and their deviations due to changes in the freezing temperature, which impose an upper limit on the effective energy density obtained from the modified Friedmann equations. The deviation from the standard energy density of the radiative plasma is therefore constrained by the abundances of the Helium-4 nuclei. Secondly, an upper limit will be considered in a numerical analysis performed through the usage of the \texttt{PRyMordial} software package, with the help of which we calculate the primordial abundances of the light elements by evaluating the thermonuclear rates within the considered modified gravity framework. Finally, an MCMC analysis will validate the cosmological model with Weylian boundary contributions, imposing relevant constraints on the initial conditions of the cosmos. The methodology is implemented in the python code \texttt{genesys}, which is available on GitHub.
	\end{abstract}
	
	\maketitle
	\tableofcontents
	
	\section{Introduction}

	The Big Bang model emerged from various theoretical developments and observations of the 20th century, and proposes that the Universe originated from an infinitely hot and dense region of space-time, generically called a singularity, and in which the geometrical and physical quantities diverge \cite{Peeb1}. 

Many important scientists contributed to the early formulation of this theory, some of the most notable being G. Lema\^{i}tre, G. Gamow and R. Alpher \cite{Peeb2}. Based on the observations of the recession of galaxies, they suggested that the chemical elements were formed after a primordial thermonuclear explosion, which took place in an environment in which the temperature, pressure and density were extremely high. These insights, together with the discovery of the Cosmic Microwave Background (CMB) radiation, offered a solid theoretical foundation for the concept of the beginning of the Universe, marked chronologically as the "Big Bang". The CMB radiation originates from the decoupling of photons from baryonic matter, due to the cooling of the Universe, which occurred due to the expansion of the space \cite{Muchanov, CMB1}.
	
	Following the Big Bang, the Universe underwent a period of rapid expansion, which led to the formation of the present-day cosmos. The de Sitter type exponential growth is unexplainable by the Big Bang theory alone. Cosmological inflation, introduced by Guth \cite{Guth} explained the early type accelerated phase  by including in the cosmological models a scalar field, called inflaton, whose potential energy must briefly dominate the energy density of the Universe. With this assumption one could explain and solve the flatness, horizon,  and magnetic monopole problems, respectively. Initially, the inflation era was thought to occur so rapidly that the interactions of the scalar field with other fields were non-existent, leading to a supercooling of the Universe.  \\

Various theories were proposed within the cold inflationary regime, such as slow roll inflation \cite{L-1982}, characterized by a prolonged period of rapid expansion, and chaotic inflation \cite{AS, L-1983, L-1994, Liddle}, governed by a massive scalar field, and a high potential energy. Subsequently many other theories were developed, such as hybrid inflation, and natural inflation, together with other studies \cite{rev1, rev2, rev3, rev4, rev5, rev6, rev7}, which tried to further investigate the cold inflation models. The exit from the cold regime into the radiation-dominated Universe was explained by the reheating mechanism, which was studied extensively in \cite{Alb, DL, KL, Lyth, H2008, A, M, N1r, N2r, N3r}, and through which energy was transferred from the inflaton scalar field to the Standard Model particles, and especially radiation. This transition marked the shift from the inflationary phase to the hot Big Bang era.
	
A relatively recent field of study is provided by the warm inflation theory, which proposes that the early Universe experienced a phase characterized by thermal fluctuations. This theory is in contrast to the cold, purely quantum fluctuations typically associated with standard inflation. 

The approach developed in \cite{BF, B, H} has shown that by allowing the inflaton field to interact dynamically with the newly created particles via thermal interactions, the energy is dissipated more gradually. Therefore, the reheating mechanism is no longer necessary for the transition to a Universe filled with radiation. The interplay between the scalar field and a thermal bath of particles helps in maintaining a state of thermal equilibrium throughout the very early Universe, which both allows to sustain the inflationary expansion, and also successfully explains the observed temperature anisotropies in the CMB radiation. 

In the warm inflationary paradigm, the interrelationship between the scalar field and the thermal bath of particles (photons)  is described in a simple way by the energy balance equations
	 \begin{eqnarray}\label{1a}
	 	\dot{\rho}_{\phi }+3H\left( \rho _{\phi }+p_{\phi }\right) &=&-\Gamma \dot{%
	 		\phi}^{2} \\
	 	\dot{\rho}_{rad}+3H\left( \rho _{rad}+p_{rad}\right) &=&\Gamma \dot{\phi}%
	 	^{2},
	 \end{eqnarray}
which represent the decay of the scalar field, and the subsequent creation of radiation. In the above equations  $\Gamma $ denotes the dissipation coefficient. The evolution of the scalar field is given by a Klein-Gordon type equation
	 \begin{equation}
	 	\ddot{\phi}+3H(1+Q)\dot{\phi}+V^{\prime }(\phi )=0,
	 \end{equation}%
 where the coefficient $Q=\Gamma /3H$ denotes the ratio of the dissipation coefficient to the Hubble function $H$, which gives the expansion rate of the Universe  \cite{H}. Additionally, considering the principles of statistical mechanics, the scalar field would tend to distribute its energy to other fields, as the system would seek to evenly distribute the accessible energy. For subsequent development in this area of research see \cite{O, Moss, Zhang, W1, W2, W3, W4, W5, W6, W7, W11, W12, W28, W29, W30, W37a, W49}. For a recent review of the warm inflationary models, including a historical perspective, see \cite{Berera}. 
	 
	 As the early Universe was much smaller than it is today, the inclusion of a physical boundary must be taken into account when studying the initial conditions for the evolution of the cosmos \cite{GH, Y}. Various approaches have addressed the incorporation of boundary terms into the dynamics of the primordial Universe, and notably, these effects are believed to have a geometric origin \cite{BR1, BR2, B3}. J. S. Ridao and M. Bellini \cite{BR1,BR2} defined the variation of the Ricci curvature tensor  in the Hilbert-Einstein variational principle, $\delta S=\delta \int{g^{\alpha \beta }R_{\alpha \beta}\sqrt{-g}d^4x}$ as $\delta R_{\alpha \beta}\equiv \nabla_{\mu} W^{\mu}=\phi(x^{\alpha})$, where the quantity $W$ represents a geometric tetra-vector, while $\phi$ is an arbitrary scalar field that accounts for the back-reaction effects caused by boundary contributions. The field produced by this tetra-vector remains invariant under the gauge transformation $\delta \tilde W_\alpha = \delta W_\alpha - \Lambda g_{\alpha \beta }$. Consequently, the Einstein tensor is expressed as $\tilde G_{\alpha \beta}=G_{\alpha \beta }-\Lambda g_{\alpha \beta}$, leading to the field equations $G_{\alpha \beta }+\Lambda g_{\alpha \beta}=\kappa T_{\alpha \beta}$. 

In order to construct geometric theories of gravitation, it is important to consider an Einstein-Hilbert Lagrangian that remains invariant under conformal transformations. This approach  requires to adopt a non-metric, Weyl type  geometry \cite{Weyl1, Weyl2, Weyl3}

In Weyl geometry, non-metricity naturally emerges and is characterized by the Weyl vector $\omega_{\mu}$. The inclusion of the vector field  $\omega_\mu$ in the mathematical formulation of relativity leads to an extension of the gravitational action within the Weylian framework \cite{Gh0, Gh0a, fQ1,W1a,W2a,fQ2,W5a,W6a,W3a,W4a,W7a,W8a,W9a,W10a,DLS,Gh1, fQ5,fQ4}.  
A particular case of the Weyl geometry, the Weyl integrable geometry, is obtained  when the Weyl vector is represented in a pure gauge form.
	 
The existence of Weyl geometric type boundary terms in the framework of gravitational theories was already assumed in \cite{BR1, BR2}.  In the study \cite{warmweyl} the Einstein field equations of general relativity were extended by considering the existence of a Weyl type boundary. Thus, from the gravitational action in the presence of nonvanishing boundary terms one obtains a set of gravitational field equations that include new terms, of purely geometrical origin, generated by the non-metric nature of the boundary, and depending on the Weyl vector, and its covariant derivatives. In order to obtain a consistent description of the evolution of the very early Universe the presence of a dissipative scalar field was also assumed. The generalized Friedmann and Klein-Gordon equations were obtained for a Friedmann-Lemaitre-Robertson-Walker cosmological metric, under the assumption that the geometry of the boundary is Weyl integrable. Thus, the contribution of the boundary is introduced in the field equations via a second scalar field. 

This approach gives rise to three possible cosmological scenarios  for the evolution of the early Universe, obtained from an appropriate splitting of the generalized conservation equation.  In the first one the equation of motion for the boundary contribution is conserved, and both scalar fields contribute to the creation of matter. In the second one, the scalar field decouples from the field generated by the Weyl vector, and therefore the creation of radiation will be possible due to the decay of the scalar field only. Finally, a third scenario was also considered, in which only the Weyl vector drives the creation of the radiation fluid. The physical results essentially depend on the form and properties of the dissipative scalar field potential $V(\phi)$. Several forms of $V$ were considered in \cite{warmweyl}, given by $V=0$, $V\propto \phi^2$, $V= \alpha \phi^2+\beta \phi ^4$, and $V=V_0\exp \left(-\alpha \phi\right)$, respectively. 
	 
In the present work we will extend the cosmological approach considered in \cite{warmweyl} to the post-inflationary era, by considering the roles a dissipative scalar field and the boundary scalar may have played during the Big Bang Nucleosynthesis (BBN) era. The cosmological parameters of the warm inflationary models introduced in \cite{warmweyl}, namely, the coefficients in the potentials as well as the dissipation function,  can be constrained by studying the abundances of primordial nuclei - hydrogen, helium, and lithium, predicted by the Big Bang Nucleosynthesis theory. 

The period of element formation began when the temperature of the Universe dropped from beyond $10^{16}\; \rm GeV$ to $1\; \rm MeV$, the temperatures at which stable nuclei are formed as the particle reaction rate became smaller than the expansion rate of the Universe \cite{Kolb}.  Big Bang Nucleosynthesis occurred within the first three minutes of the existence of the Universe.  The existence in the very early Universe of a small content of nucleons allowed for the production of light nuclei, including hydrogen, the most common element in the Universe, helium, the second  most common element, and an isotope of lithium, $^7$Li respectively. The BBN processes occurred during a phase of still rapid expansion of the Universe, and they ended when the nuclear reactions have stopped,  due to the significant decrease of the matter densities and temperatures,  caused by the expansion of the Universe. The BBN theory was extensively studied in \cite{Benstein,Copi,Torres,Sepico, Steigman-2004, Lambiase-2005, Fields, Steigman-2007, Steigman-2008, Steigman-2012, Lambiase-2012, Cybrut, dof, Capozziello, F,fRT,obs,Hsyu,Fields1,  Barrow,Aver,Hogas, yeh,Asim, Asim1,Sultan,tau, fQ, Kho, Park,Taluk, Mis,Giri, Akar}.

The standard model of cosmology is closely related to the standard model of particle physics. Cosmological models rely on many important results of the particle physics, like the existence of three species of neutrinos. In the early Universe the formation of the light nuclei was determined by the interplay  of several physical parameter including the temperature, the expansion rate, the neutrino content,  the nucleon density,  and the neutrino-antineutrino asymmetry, respectively \cite{Kolb}. Therefore, BBN represents a powerful tool for the investigation and for the testing of the cosmological models, and of the physical parameters that determined  the formation of the primordial elements. 

Moreover, BBN offers a powerful framework for testing modified gravity theories.  In \cite{Barrow}  the BBN data were used to obtain some constraints on the Barrow entropy, an extended entropy relation that is obtained by incorporating quantum gravitational effects in the black hole thermodynamics. The parameters of the bimetric theory of gravity were constrained in \cite{Hogas}. The bimetric theory of gravity is a observationally viable, ghost free extension of general relativity, containing both a massive and a massless graviton. Constraints obtained from BBN on a number of Several higher-order modified gravity theories, obtained by using the quadratic-curvature Gauss-Bonnet $G$ term, and the cubic-curvature combination, including the $f(G)$ Gauss-Bonnet and the $f(P)$ cubic gravities were investigated in \cite{Asim}.  These models satisfy the BBN constraints, and thus can be considered as viable alternative scenarios to standard cosmology. In \cite{Asim1} the $ f \left(T ,T_G\right)$ modified gravity theory, were $T$ denotes the torsion scalar, and $T_G$ denotes the teleparallel equivalent of the Gauss–Bonnet term, was investigate from the point of view of satisfying  the BBN constraints. The presence of the extra torsion terms leads to deviations of the freeze-out temperature with respect to standard cosmology. For a large range  of model parameters the BBN constraints are satisfied by a particular logarithmic model.  The formation of light elements in the early Universe  in the Einstein-Aether and modified Horava-Lifshitz theories of gravity were investigated in \cite{Sultan}, where it was found that some models in these theories satisfy the constraints of the Big Bang nucleosynthesis. The Constraints arising from BBN on several classes of $f(Q)$ gravity models were obtained in \cite{fQ}, where it was proven that $f(Q)$ models can easily satisfy the BBN constraints, which  confirms the viability of the theory as compared to other modified gravity theories. The bumblebee model, a vector-tensor theory of gravitation, was constrained by using BBN data in  \cite{Kho}, where some restrictions on the vacuum expectation value of the bumblebee timelike vector field were obtained. 

In \cite{Park}  the BBN cosmological results were used to impose  constraints on the free parameters of the extended Starobinsky model. Non-standard cosmological models cannot generally satisfy the  BBN constraints, but for some viable models one can find an upper limit imposed by the BBN constraints. The role of the BBN constraints in various modified gravity theories, like the $f(R)$ gravity scalarons \cite{Taluk}, the $f(T,T)$ gravity theory \cite{Mis}, varying gravitational constant theories \cite{Giri}, or the quadratic energy–momentum squared gravity \cite{Akar}, was also investigated in detail. Cosmological N-body simulations with standard $\Lambda$CDM cosmology in the presence of a scale-dependent primordial non-Gaussianity were performed in \cite{Verde}.The considered models ensure consistency with current Cosmic Microwave Background bounds on primordial non-Gaussianity, also allowing for large effects of the non-Gaussian statistics on the properties of non-linear structure formation. It was also pointed out that these models might mimic the effects of Warm Dark Matter. An updated analysis of cosmological models with very low reheating scenarios was considered in \cite{Pis}, by including a more precise computation of neutrino distribution functions, by using the latest datasets from cosmological surveys. A joint analysis combining constraints from Big Bang Nucleosynthesis, the Cosmic Microwave Background, and galaxy surveys allows to obtain a lower bound on the reheating temperature of  $T_{RH}>5.96$ MeV at a 95\% confidence level, representing the most stringent constraint to date.

Thus, BBN offers the possibility of in depth testing of the modified gravity theories, as well as a powerful check of their viability and relevance. In this context, in the present paper we consider the implications on the Big Bang Nucleosynthes of the theory of gravity proposed in \cite{warmweyl}, in which a boundary contribution  significantly affects the dynamic evolution of our Universe. From a cosmological point of view we thus investigate the gravitational theory with Weylian boundary in the post-inflationary era, after the initial decay of the dissipative scalar field and of the Weyl vector generated the matter content of the Universe. 

We will constrain the theoretical models with the observational determinations  of the abundance of Helium-4, given by the primordial mass fraction estimate, whose deviation is induced by the correlated deviation of the freeze-out temperature, an approach developed in \cite{Capozziello, Lambiase-2012}. The freeze-out temperature represents the temperature at which the interconversion of neutrons and protons through weak interactions ceases. The current estimation of the primordial mass fraction and its deviation are discussed in \cite{Aver, Hsyu, Fields1, pdg}. Therefore, the contribution to the energy density coming from the dissipative inflation field and the Weyl contribution will be constrained by an upper limit imposed by the $\delta Y_p/Y_p$ fraction.

	 Additionally, we will introduce the effective energy density along with its constrained contribution in a python language software package called \texttt{PRyMordial} \cite{primordial}, so that the three cosmological models can integrated in the BBN framework. Within the library, the light nuclei abundances are calculated  by computing thermonuclear rates available in NACRE II database \cite{nacre2}, using the Friedmann equations for describing the plasma. 
	 
	 The present paper is organized as follows. The warm inflation scenario and the derivation of the general Friedmann equations from the field equations in the presence of a Weylian boundary  are reviewed in Section \ref{dyn}. Fundamental concepts of Big Bang Nucleosynthesis theory are presented in Section \ref{bbn}, along with a thermal history of the Universe, where particle interactions play a significant role. Additionally, the BBN constraining method is described along with the derivation of the upper limit for the effective energy density. The dimensionless form of the generalized Friedmann equations is presented in Section~\ref{numerical1}, where the first constraints on the energy density and the parameter of the equation of state of the effective dark energy are also prseented. The numerical analysis is described in Section \ref{numerical}, where we integrated the models into the \texttt{PRyMordial} framework by computing the additional energy density and pressure of the plasma. Moreover, the validity of our theoretical models is analysed through an MCMC procedure that evaluates the posterior distributions of the abundances of light nuclei within both the standard BBN and the theoretical models. The constraints on the scalar field potentials of the three cosmological models obtained for the statistical analysis and comparison with the BBN data are described in Section~\ref{numerical2}. Finally, the conclusions concerning our investigations, and relevant discussions are presented in Section~\ref{concl}.
	 
\section{The dynamics of the early Universe in the presence of a Weylian boundary}\label{dyn}
	
	In the present Section, we will introduce the theoretical scenario, within which a Weylian boundary influence plays an essential role in the dynamics of the early Universe. The variation of the gravitational action is extended by keeping the boundary terms defined in the Weyl geometry. Consequently, the obtained gravitational field equations also include the effects of the Weyl vector, coming from the boundary terms, in the early evolution of the Universe. The interaction of the scalar field with a thermal bath of particles, resembling the matter component, is considered as a source of energy dissipation. A generalization  of the Friedmann equations leads to the formulation of three distinct cosmological evolutionary  models of  the Universe, in the presence of the scalar field and the Weyl vector. These three models correspond to a Universe in which only the scalar field drives the cosmological evolution,  and one in which the Weyl effects are driving the expansion through the dynamics of the Weyl vector.
	
	\subsection{Brief introduction to Weyl Geometry}
	
	Weyl geometry is defined through the classes of equivalence $\left(g_{\alpha \beta}, \right)$ of the metric $g_{\alpha \beta}$ and Weyl gauge field $\omega _\mu$, related by a Weyl gauge transformation given by\cite{Gh1},
	\begin{eqnarray}\label{WS}
		\text{(i)} \quad \hat g_{\mu\nu}&=&\Sigma^n \,g_{\mu\nu},\qquad \sqrt{-\hat g%
		} =\Sigma^{2 n} \sqrt{-g},  \notag \\
		\text{(ii)} \quad \hat \phi &=& \Sigma^{-n/2} \phi, \qquad
		\;\;\hat\psi=\Sigma^{-3n/4}\,\psi,
	\end{eqnarray}
	where $g=\det g_{\mu\nu}$, and $\Sigma(x)>0, \forall x$. In the following $n$ is assumed as $n=1$. Here  $\phi$ and $\psi$ represent the bosonic and fermionic fields. 
	The previous equations, together with the transformation rule of the Weyl vector $\omega_\mu $,  
\begin{equation}\label{WGS}
		\hat{\omega}_{\mu }=\omega _{\mu }-\frac{1}{\alpha }\,\partial _{\mu }\ln
		\Sigma ,
	\end{equation}%
where $\alpha $ denotes the Weyl gauge coupling parameter, give the necessary  conditions  for constructing conformally invariant theories  in the Weyl geometry.
	In this geometry non-metricity arises naturally via the nonmetric nature of the covariant derivative of the metric tensor
	\begin{equation}
		\tilde{\nabla}_{\mu }g_{\alpha \beta }=-\alpha \omega _{\mu }g_{\alpha \beta}.
	\end{equation}

	From the above relation we obtain the expression of the connection in Weyl geometry as given by
	\begin{equation}\label{conn}
		\tilde{\Gamma}_{\mu \nu }^{\lambda } =\Gamma _{\mu \nu }^{\lambda }+\frac{1%
		}{2}\alpha \big[\delta _{\mu }^{\lambda }\,\,\omega _{\nu }+\delta _{\nu
		}^{\lambda }\,\,\omega _{\mu }-g_{\mu \nu }\,\omega ^{\lambda }\big]
	\end{equation}%
where by $\Gamma _{\mu \nu }^{\lambda }$ we have denoted  the Levi-Civita connection of Riemannian geometry. 
	
	The field strength $F_{\mu\nu}$ of the Weyl vector $\omega_\mu$, is defined as
	\begin{equation}  \label{W}
		\tilde{F}_{\mu\nu} = \tilde{\nabla}_{\mu} \omega_{\nu} - \tilde{\nabla}
		_{\nu} \omega_{\mu} = \nabla _{\mu}\omega _\nu-\nabla _\nu \omega _\mu.
	\end{equation}
	
	Using the Weyl connection, $\tilde\Gamma$, we can obtain the Riemann and the Ricci curvatures as well as the curvature scalar in Weyl geometry as follows
	\begin{eqnarray}
		\tilde R^\lambda_{\mu\nu\sigma}&=& \partial_\nu
		\tilde\Gamma^\lambda_{\mu\sigma} -\partial_\sigma
		\tilde\Gamma^\lambda_{\mu\nu}
		+\tilde\Gamma^\lambda_{\nu\rho}\,\tilde\Gamma_{\mu\sigma}^\rho
		-\tilde\Gamma_{\sigma\rho}^\lambda\,\tilde\Gamma_{\mu\nu}^\rho,  \notag \\
		\tilde R_{\mu\nu}&=&\tilde R^\lambda_{\mu\lambda\sigma}, \tilde
		R=g^{\mu\sigma}\,\tilde R_{\mu\sigma}.
	\end{eqnarray}
	
	In terms of the Riemannian geometric quantities and the Weyl vector the Ricci tensor and the Ricci scalar can be expressed as
	\begin{eqnarray}  \label{tRmunu}
		\tilde R_{\mu\nu}&=&R_{\mu\nu} +\frac {1}{2}\alpha \left(\nabla_\mu \omega
		_\nu-3\,\nabla_\nu \omega _\mu - g_{\mu\nu}\,\nabla_\lambda \omega
		^\lambda\right)  \notag \\
		&&+\frac{1}{2} \alpha ^2 (\omega _\mu \omega _\nu -g_{\mu\nu}\,\omega
		_\lambda \omega ^\lambda),
	\end{eqnarray}
	\begin{eqnarray}
		&&\tilde R= R-3 \,\alpha\,\nabla_\lambda\omega ^\lambda-\frac{3
		}{2}\alpha ^2 \omega _\lambda \omega ^\lambda,
	\end{eqnarray}
	where $R_{\mu \nu}$ and $R$ represent the Ricci tensor and the Ricci scalar obtained with the Levi Civita connection,  respectively, with the covariant derivative defined conventionally as $\nabla_\mu\omega
	^\lambda=\partial_\mu \omega ^\lambda+\Gamma^\lambda_{\mu\rho}\,\omega ^\rho$.

The Weylian Ricci tensor is antisymmetric, and it satisfies the relation
\be
\tilde R_{\mu\nu}-\tilde R_{\nu\mu}=2 \alpha F_{\mu\nu}.
\ee
Moreover, $\tilde R$ transforms covariantly under the conformal transformations as $\hat{\tilde R}=(1/\Sigma^n)\,\tilde R$.

\subsection{Action and field equations in a Universe with a Weylian boundary}

The starting point of our approach is the Hilbert-Einstein action, and variational principle, whose purely gravitational sector is written down in its standard form 
in Riemann geometry,  and given by
\begin{equation}\label{act1}
\delta S_{g}=-\delta \frac{1}{2\kappa^2}\int_{\Omega }{R(g)\sqrt{-g}d^{4}x}=0,
\end{equation}
where $\kappa ^2=8\pi G/c^4$. By varying the action with respect to the metric we obtain
\bea\label{25}
\delta S_g&=&-\frac{1}{2\kappa ^2}\int_\Omega\left(R_{\mu \nu}-\frac{1}{2}Rg_{\mu \nu}\right)\delta g^{\mu\nu}\sqrt{-g}d^4x\nonumber\\
&&-\frac{1}{2\kappa ^2}\int_\Omega {\left(g^{\mu
		\nu }\nabla _{\lambda }\delta \Gamma _{\mu \nu }^{\lambda }-g^{\mu \lambda
	}\nabla _{\lambda }\delta \Gamma _{\mu \sigma }^{\sigma }\right)\sqrt{-g}d^4x}.\nonumber\\
\eea

The basic idea of \cite{warmweyl} is to replace in the boundary term the ordinary Riemannian covariant derivative with the Weylian one, so that $\nabla _\lambda \rightarrow \tilde{\nabla}_\lambda$, where $\tilde{\nabla}_\lambda$ is constructed with the help of the Weyl connection (\ref{conn}). Thus, a new degree of freedom, the Weyl vector, is naturally introduced in the theoretical formalism describing the gravitational interaction. Hence, for the variation of the boundary term we obtain the relation     
\begin{eqnarray}\label{R}
	 	&&g^{\mu\nu}\delta \tilde{R}_{\mu\nu} = g^{\mu \nu }\tilde{\nabla}_{\lambda }\delta \tilde{\Gamma}%
	 	_{\mu \nu }^{\lambda }-g^{\mu \lambda }\tilde{\nabla}_{\lambda }\delta
	 	\tilde{\Gamma}_{\mu \sigma }^{\sigma }=\frac{\alpha }{2}\times \nonumber\\
	 	&& \Big[\Big(2\tilde{%
	 		\nabla}_{\gamma }\omega _{\rho }-g_{\gamma \rho }\tilde{\nabla}_{\lambda
	 	}\omega ^{\lambda }
	 	-g^{\mu \lambda }g_{\gamma \rho }\tilde{\nabla}_{\lambda
	 	}\omega _{\mu }\Big)
	 	+\alpha \omega ^{\lambda }\omega
	 	_{\lambda }g_{\gamma \rho }\Big]\delta g^{\rho \gamma },\nonumber\\
	 \end{eqnarray}
which leads to the generalized Einstein  equations with Weylian boundary terms, which are given by \cite{warmweyl}
\bea
&&R_{\mu \nu}-\frac{1}{2}Rg_{\mu \nu}+\alpha \left(\nabla _\mu \omega _\nu-g_{\mu \nu}\nabla _\lambda \omega ^\lambda\right)\nonumber\\
&&-\alpha ^2\left(\omega _\mu \omega _\nu+\frac{1}{2}\omega ^2 g_{\mu \nu}\right)= \kappa^2 T_{\mu \nu}^{(m)}+ T^{(\phi)}_{\mu \nu},
\eea
where
\be
T_{\mu \nu}^{(m)} = \frac{2}{\sqrt{-g}}\frac{\partial (\sqrt{-g}\mathcal{L}%
_{m})}{\partial g^{\mu \nu }},
\ee
is the energy-momentum tensor of the ordinary matter, and we have also introduced a cosmological scalar field $\phi$ with energy-momentum tensor denoted by $T^{(\phi)}_{\mu \nu}$.

In order to simplify the mathematical formalism, and to gain some simpler insights into the possible role of the boundary terms in the cosmological evolution, we assume now that the Weyl vector is given by the gradient of a scalar function $\psi$, with $\omega _\mu =\nabla _\mu \psi$. Thus we assume that the boundary of the Universe is described by the integrable Weyl geometry. As a second approximation we assume that the Weylian geometric effects are small, and therefore one can ignore the terms $\omega ^2/2$, and $\nabla _\mu \nabla _\nu \psi$ in the field equations. Since $\nabla _\lambda \omega ^\lambda=\nabla _\lambda \nabla ^\lambda \psi=\Box \psi$, the simplified field equations describing the evolution of a Universe with a Weyl integrable boundary are given by 
	\begin{eqnarray}
		 R_{\mu \nu}-\frac{1}{2}Rg_{\mu \nu} - \alpha \Bigg( \partial_{\mu}\psi \partial_{\nu}\psi +\alpha g_{\mu \nu}\Box \psi \Bigg)= \kappa^2 T_{\mu \nu}^{(m)}+ T^{(\phi)}_{\mu \nu}.\nonumber\\
	\end{eqnarray}
After rescaling the Weyl scalar according to $\psi \rightarrow \alpha \psi$, and redefining $\alpha$ as $\alpha ^3 \rightarrow \alpha ^2/2$, the gravitational field equations take the form
	\begin{eqnarray}\label{fe}
		 R_{\mu \nu}-\frac{1}{2}Rg_{\mu \nu} - \frac{\alpha ^2}{2} \Bigg( \partial_{\mu}\psi \partial_{\nu}\psi +g_{\mu \nu}\Box \psi \Bigg)= \kappa^2 T_{\mu \nu}^{(m)}+ T^{(\phi)}_{\mu \nu}.\nonumber\\
	\end{eqnarray}

The energy-momentum tensor for the matter is given by
\begin{equation}
	T_{\mu \nu}^{(m)}=\left(\rho_m+p_m\right)u_\mu u_\nu -p_m g_{\mu \nu},
\end{equation}
where $\rho_m$ and $p_m$ are the energy density and pressure of the ordinary matter.  

The Lagrange density of the dissipative scalar field is represented in the form  \cite{dis}
\begin{equation}
\mathcal{L}_{\phi }= e^{\Gamma(\phi)} \left[ \frac{1}{2}g^{\mu \nu } \frac{%
\partial \phi }{ \partial x^{\mu }}\frac{\partial \phi }{\partial x^{\nu }}%
-V\left( \phi\right) \right]\sqrt{-g},  \label{L}
\end{equation}
where $\Gamma$ is the dissipation function. In the following we will assume for $\Gamma$ the simple expression $\Gamma = \beta \phi$, where $\beta $ is a constant.

Hence, the energy-momentum tensor of the dissipative scalar field, which follows from (\ref{L}), has the form
\begin{eqnarray}\label{tdis}
\hspace{-0.5cm}T_{\mu \nu }^{(\phi)}
= e^{\Gamma (\phi)} \left[\partial_{\mu}\phi \partial_{\nu}\phi-g_{\mu \nu }\left( \frac{1}{2}g^{\sigma \rho }\partial_{\sigma}\phi
\partial_{\rho}\phi-V(\phi)\right)\right].\nonumber\\
\end{eqnarray}
It can be also represented in an equivalent form, similar to the one describing ordinary matter, given by
\bea
	T_{\alpha \beta} ^{(\phi)}= (p_{\phi}+\rho_{\phi})u_{\alpha}u_{\beta}- p_{\phi}g_{\alpha \beta}, \nonumber\\
\eea
where
\begin{eqnarray}\label{p_phi}
	\rho_{\phi} &=& e^{\Gamma (\phi)}\left( \frac{1}{2}g^{\mu \nu }\phi _{,\mu }\phi
	_{,\nu }+V(\phi) \right), \\ 
p_{\phi}& =& e^{\Gamma (\phi)}\left( \frac{1}{2}g^{\mu \nu }\phi _{,\mu }\phi _{,\nu
	}-V(\phi) \right), \\
	u_{\mu}^{(\phi)} &=& \frac{\phi_{,\mu}}{\sqrt{g^{\alpha \beta }\phi _{,\alpha }\phi_{,\beta}}},
\end{eqnarray}
where $\rho_\phi$, $p_\phi$ and $u_{\mu}^{(\phi)}$ are the energy density, pressure and velocity of the dissipative scalar field. $V(\phi)$ denotes the self-interaction potential of the field.

The generalized Klein-Gordon equation giving the evolution of the dissipative scalar field is obtained, for $\Gamma =\beta \phi$, as \cite{dis}
\begin{equation}\label{eom}
	\Box\phi + \frac{dV(\phi)}{d\phi} + \beta \left[g^{\mu \nu}\partial_{\mu}\phi \partial_{\nu}\phi+V(\phi)\right] = 0.
\end{equation}

\subsection{Generalized Friedmann equations in the presence of a boundary} \label{models} 
	
	The early evolution of the Universe is considered in the  flat, homogeneous, and isotropic standard Friedman-Lema\^{i}tre-Robertson-Walker (FLRW) metric, given by
	\begin{equation}
		ds^2 = c^2dt^2 - a^2(t)(dx^2 + dy^2 + dz^2),
	\end{equation}
where $a(t)$ is the scale factor. An important observational quantity, the Hubble function $H(t)$, describing the expansion rate of the Universe, is defined as $H(t)=\dot{a}(t)/a(t)$, where a dot denotes the derivative with respect to the cosmological time $t$. 

	In the FLRW metric, the equation of motion of the dissipative scalar field, or the generalized Klein-Gordon equation,  is 
	\begin{equation}\label{KG}
		\ddot{\phi} + 3H\dot{\phi} + V^{\prime}(\phi) = -\beta \dot{\phi}^2 - \beta V(\phi).
	\end{equation}

		The relationship between the expansion rate, and the dynamics of the scalar field and matter is described by the Friedmann equations. In the present model, for a Universe with a Weylian boundary, the first and the second Friedmann equation are derived from the field equations Eq.~(\ref{fe}), and they are obtained as  
	\begin{eqnarray}\label{F1}
	\hspace{-0.5cm}3H^2 &=& 8\pi G \rho_m+e^{\beta \phi}\left(\frac{\dot{\phi}^2}{2}+V(\phi)\right) \notag \\
		\hspace{-0.5cm}&+&\frac{\alpha^2}{2}\left(\ddot{\psi}+3H\dot{\psi}+\dot{\psi}^2\right)=8\pi G \rho_m+\rho_{eff}, 
\eea
\bea\label{F2}
	\hspace{-0.2cm}	2\dot{H} + 3H^2 &=& -\frac{8\pi G}{c^2}p_m- e^{\beta \phi}\left(\frac{\dot{\phi}^2}{2}-V(\phi)\right)\notag \\
	\hspace{-0.2cm}	&+&\frac{\alpha^2}{2}\left(\ddot{\psi}+3H\dot{\psi}\right)=-\frac{8\pi G}{c^2}p_m-p_{eff}.
	\end{eqnarray}
	
	Along with the previous set of relations, the energy balance equations, which follow from the divergence of the total energy-momentum tensor of matter, scalar field, and Weyl vector contribution,  offer a complete description of the dynamic interaction between the cosmic constituents. From Eqs.~(\ref{F1}) and (\ref{F2}) we obtain the global conservation equation
	\begin{eqnarray} \label{cons_eq}
		&&8\pi G\left[\dot{\rho}_m+3H\left(\rho_m+\frac{p_m}{c^2}\right)\right]-\beta \frac{\dot{\phi}^3}{2}e^{\beta \phi}\notag \\
		&&+\frac{\alpha ^2}{2}\frac{d}{dt}\left(\ddot{\psi}+3H\dot{\psi}+\dot{\psi}^2\right)+\frac{3\alpha ^2}{2}H\dot{\psi}^2=0.
	\end{eqnarray}
	
	Several models that account for the description of the boundary contribution on the cosmological dynamics were developed in \cite{warmweyl}. The models were used for the study of warm inflation, including the radiation production. In the following we will investigate the effects of these models on the BBN processes. We present, for the sake of completeness,  the corresponding cosmological models in the following. They are easily obtained by splitting the general conservation equation in different ways.
	
\paragraph{Model I.} The first cosmological model is obtained by splitting the conservation equation Eq.~(\ref{cons_eq}),  so that the Weyl boundary contribution is conserved. Consequently, the energy balance equation is split into two equivalent equation, 
	\begin{equation}\label{M2}
		\ddot{\psi}+3H\dot{\psi}+\dot{\psi}^2=\Lambda={\rm constant}, 
	\end{equation}
	\begin{equation}\label{M4}
		8\pi G\left[\dot{\rho}_m+3H\left(\rho_m+\frac{p_m}{c^2}\right)\right]=\beta \frac{\dot{\phi}^3}{2}e^{\beta \phi}-\frac{3\alpha ^2}{2}H\dot{\psi}^2.
	\end{equation}
	

	\paragraph{Model II.} The second cosmological model is constructed by assuming that the scalar fields $\phi$ and $\psi$ decouple in the general conservation equation, and the radiation creation process is determined only by the scalar field,
	\begin{equation}\label{M6}
		\frac{\alpha ^2}{2}\frac{d}{dt}\left(\ddot{\psi}+3H\dot{\psi}+\dot{\psi}^2\right)+\frac{3\alpha ^2}{2}H\dot{\psi}^2=0,
	\end{equation}
	\begin{equation}\label{M8}
		8\pi G\left[\dot{\rho}_m+3H\left(\rho_m+\frac{p_m}{c^2}\right)\right]=\beta \frac{\dot{\phi}^3}{2}e^{\beta \phi}.
	\end{equation}
	
	This scenario allows the creation of matter to be driven only by the decay of the scalar field $\phi$, which is described by Eq.~(\ref{M8}). 
	
\paragraph{Model III.} A third splitting of the general energy balance equation into the following relations gives a third  cosmological model
	
	\begin{equation}\label{M10}
		\beta \frac{\dot{\phi}^3}{2}e^{\beta \phi}=\frac{3\alpha ^2}{2}\frac{d}{dt}\left(\ddot{\psi}+3H\dot{\psi}+\dot{\psi}^2\right)+\frac{3\alpha ^2}{2}H\dot{\psi}^2,
	\end{equation}
	\begin{equation}\label{M12}
		8\pi G\left[\dot{\rho}_m+3H\left(\rho_m+\frac{p_m}{c^2}\right)\right]=\alpha ^2\frac{d}{dt}\left(\ddot{\psi}+3H\dot{\psi}+\dot{\psi}^2\right).
	\end{equation}
	
	This last model considers that the creation of matter via the nonconservation of the matter energy-momentum tensor is a direct effect of the evolution of the Weyl vector field. The dynamic relationship between the scalar and the Weyl fields is described  by Eq.~(\ref{M12}). From Eq.~(\ref{M12}) one can infer that the decay of the scalar field and the evolution of the boundary term are closely correlated.

\section{Brief review of standard BBN theory} \label{bbn}
	
	In the following section, we will introduce a brief description of the particle interactions that took place in the early Universe. The dynamic evolution of photons, electrons and positrons, neutrinos and baryons under the influence of temperature eventually led to the synthesis of light nuclei, deuterium, tritium, helium-3, helium-4 and lithium-7.

	Additionally, the primordial mass fraction estimate will provide useful constraints on evaluating the contribution of the effective energy density to the creation of light nuclei. As the freeze-out temperature deviation is correlated to the Helium-4 abundance, the method analyses the impact of Helium-4 deviations into the plasma evolution for physics beyond Standard Model.

\subsection{Basic physical processes of the BBN}

	In the early Universe  the particle reaction rate is determined by the interconversion rate of neutrons and protons, which is possible through weak interactions and $\beta$ decay,
	 \begin{eqnarray}
	 	n+\nu_e&\rightarrow& p+e^-, \\
	 	n+e^- &\rightarrow& p + \overline{\nu_e}, \\
	 	n &\rightarrow& p + e^- + \overline{\nu_e}.
	 \end{eqnarray}
	 
	 As the temperature drops further, the weak interactions are no longer permitted energetically, thus the particles exit thermal equilibrium, causing a freeze of the $n_n/n_p$ ratio. A short period of time will pass between the "freeze-out", which occurs around $T_f\sim 0.5-0.6\; \rm MeV$, and the beginning of the nucleosynthesis at $T_n\sim0.1\; \rm MeV$.
	 
	 By entering the period in which the reaction rate becomes smaller than the expansion rate of the Universe, the thermal equilibrium achieved through the interconversion of neutrons and protons is disturbed, and therefore the neutron-to-proton ratio,
	 \begin{equation}
	 	\frac{n_n}{n_p}=\frac{x_n}{x_p}=e^{-Q/T},
	 \end{equation}
	 will be frozen at $e^{-Q/T_f}\approx 1/6$, where $Q= m_n-m_p = 1.29 \times 10^{-3}\; \rm GeV$ is the neutron-to-proton mass difference and $T_f$ is the freeze-out temperature. From the theory of weak interactions, the freeze-out temperature is estimated around $T_f \approx 0.6\; \rm MeV$ \cite{Barrow}. 
	 
	 The neutron decay decreases the ratio until all neutrons are bound to nuclei. These light elements are deuterium, $^3$H, and $^3$He which contribute further to the synthesis of $^4$He and $^4$Li.
	 \begin{eqnarray}
	 	n + p &\rightarrow& ^2H + \gamma, \notag \\ \notag
	 	d + p &\rightarrow& ^3He + \gamma, \quad d + d \rightarrow ^3He + n + \gamma, \\ 
	 	d + n &\rightarrow& ^3H, \notag \\ \notag
	 	^3He + n &\rightarrow& ^4He + \gamma, \quad ^3He + d \rightarrow ^4He + p, \\
	 	^3H + p &\rightarrow& ^4He + \gamma,\quad ^3H + d \rightarrow ^4He + n, \notag \\
	 	^4He + ^3H &\rightarrow& ^7Li + \gamma, \quad ^4He + ^3He \rightarrow ^7Li + \gamma.
	 \end{eqnarray}
	 
	 Almost all neutrons contributed to the formation of helium-4, as 25\% of all mass in the universe is in the form of $^4$He. The nucleus of helium has two neutrons, therefore the mass fraction of $^4$He can be expressed in terms of the neutron and proton density,
	 \begin{equation}
	 	Y_p=\frac{4(n_n/2)}{n_n+n_p}=\frac{2(n/p)}{1+(n/p)}.
	 \end{equation}
	 The abundance of D, $^3$He and $^7$Li are given by numerical best fits of the baryon number density. 

	\subsection{Particle interactions in the early Universe}
	
	The interactions of particles through Coulombian forces, scatterings and energy-momentum exchange contribute to the reaction rate $\Gamma(t)$, which exceeds the expansion rate of the Universe during the first moments, thus maintaining thermal equilibrium among photons, baryons and leptons, which were created in the primordial Universe. 
	When the reaction rate of a particle species drops below the expansion rate of the Universe, $\Gamma(t)\ll H(t)$, for a temperature $T>m_e$, the particle decouples and its distribution function freezes in a form that can be traced at any time following the decoupling, $t>t_D$.
	
	Notably, during the radiation-dominated phase which follows the cosmic inflation, at a time $t<t_D$, the scale factor of the Universe $a(t)\propto t^{1/2}$ and as the particles existent at that time are relativistic, the following relation can be established,
	\begin{equation}
		\left(\frac{\dot{a}}{a}\right)^2 = H(t)^2 = \frac{1}{4t^2}=\frac{8\pi G}{3}\rho.
	\end{equation}
	The energy density of the relativistic particles is defined as $\rho = \rho_{bosons}+\rho_{fermions} = g_*(\pi^2/30)T^4$ \cite{dof}. 
	
	 The radiation fluid of the early Universe consisted of photons and $e^\pm$ pairs, which co-existed in equilibrium, as well as neutrinos which come in three flavours - $\nu_e, \; \nu_\mu$ and $\nu_\tau$. Notably, the abundance of baryonic matter is determined by computing the number of baryons relative to the number of the CMB photons,
	\begin{equation}
		\eta = n_B/n_\gamma \approx 10^{-10},
	\end{equation}
	a quantity which remains constant from that time to our day \cite{Steigman-2004}.
	
	 The interconversion of photons into $e^\pm$ pairs is the process through which the thermodynamic equilibrium is maintained, which leads to a conservation of chemical potential $\mu_e^- +\mu_e^+=0$, as $\mu_\gamma=0$. Prior to $e^\pm$ annihilation, the energy density of radiation can be expressed as
	 \begin{equation}
	 	\rho_{r} = \rho_{\gamma} + \rho_e + 3 \rho_{\nu} = \frac{43}{8} \rho_{\gamma}.
	 \end{equation}
	 
	 Neutrinos are known to be electrically neutral so that they are not coupled to photons. As temperature reaches a value of $T\simeq 1\; \rm MeV$, the reaction rate of neutrino interactions falls short the expansion rate, $\Gamma_{\nu}< H(t)$, thus neutrinos decouple from matter and as there are relativistic particles, a relic background of neutrinos is formed. In SBBN, this decoupling is thought to occur before the $e^\pm$ annihilation, and the temperatures of the photon and neutrino background satisfy the relation
	 $T_\gamma / T_\nu = (11/4)^{1/3}$. 
	 
	 In a non-standard framework, the effective energy density can be expressed in terms of the standard and extra energy density contribution,
	 \begin{equation}
	 	\rho_{r}^{(eff)} = \rho_{r} + \rho_{X},
	 \end{equation}
	 with $X$ denoting the contribution to the energy density of particle species or geometric effects beyond the Standard Model. Thus, the energy density associated can be expressed as
	 \begin{equation}
	 	\rho_{X} = \Delta N_\nu \rho_\nu = \frac{7}{8} \Delta N_\nu \rho_\gamma,
	 \end{equation}
	 where $\Delta N_\nu < 0.18$ for a 2$\sigma$ confidence level \cite{yeh}.
	 
	 The Universe is observed to be electrically neutral and as the proton-to-photon ratio is approximately $n_p/n_\gamma \approx 10^{-10}$, the electron abundance will also have a value of 
	 \begin{equation}
	 \frac{n_e^- - n_e^+}{n_\gamma} \approx 10^{-10}.
	 \end{equation}
	  Consequently, a quantity called lepton asymmetry is defined in literature, which is also a BBN parameter to which $\rm ^4He$ is most sensitive \cite{Steigman-2012},
	 \begin{equation}
	 	\sum_{\alpha}\frac{\left(n_\nu - n_{\overline{\nu}}\right)_\alpha}{n_\gamma} = \frac{\pi^3}{12 \zeta(3)}\sum_{\alpha}\left(\frac{\xi_\alpha}{\pi}+ \left(\frac{\xi_\alpha}{\pi}\right)^3\right),
	 \end{equation}
	 where the sum is calculated over all three flavours of neutrinos, ($\alpha=e,\tau,\mu$) and $\zeta(n)$ is the Riemann zeta function. A dimensionless degeneracy parameter $\xi_\alpha$ is defined as the ratio of neutrino chemical potential to the neutrino temperature, $\xi_\alpha=\mu_\alpha/T_\alpha$.

	Furthermore, as the temperature drops below $T < m_e = 0.5\; \rm MeV$, the energy carried by photons will also reach a value below the energy required for pair creation of $e^\pm$. The particle-antiparticle pairs will annihilate, leaving an excess of electrons over positrons, which matches the number of protons contained by the Universe to confer charge neutrality. The photons remain the only relativistic species left.

	\subsection{BBN constraints from the freeze-out temperature deviation} \label{method1}
	During the radiation-dominated epoch, when the BBN occurs, the particles filling the early Universe are subject to relativistic physics and their energy density is given by
	\begin{equation}
		\rho _r= \frac{\pi^2}{30} g^* T^4, \label{enden}
	\end{equation}
	where $g_*$ is the effective number of degrees of freedom, $g_* \sim 10$,
	and $T$ the temperature \cite{Capozziello}. The neutron abundance is given by the rate of conversion of protons to neutrons,
	\begin{equation}
		\Lambda(T)=\Lambda_{n+\nu_e\rightarrow p+e^-} + \Lambda_{n+e^- \rightarrow p + \overline{\nu_e}} + \Lambda_{n \rightarrow p + e^- + \overline{\nu_e}}, 
	\end{equation}
	and is given by
	\begin{equation}
		\Lambda(T)=4AT^3(4!T^2+2\times3!QT+2!Q^2),
	\end{equation}
	where $Q = m_n - m_p = 1.29 \times 10^{-3}\; \rm GeV$ is the neutron to proton mass difference and $A=1.02\times 10^{-11} \; \rm GeV^{-4}$, \cite{Barrow}. The particles are in thermal equilibrium if $1/H \ll \Lambda(T)$, where $H$ denotes the Hubble parameter, and for $1/H \gg \Lambda(T)$ the particles decouple. The phenomenon known as the freeze-out of the temperature of the Universe $T_f$ occurs at $H=\Lambda(T)| _{T=T_f}$, \cite{Kolb}. The weak interaction rate of particles  therefore becomes \cite{Lambiase-2012, Benstein}
	\begin{equation}
		\Lambda(T)\simeq qT^5 + \mathcal{O}\left(\frac{Q}{T}\right),
	\end{equation}
	where $q= 4A4!=9.6 \times 10^{-10}\; \rm GeV^{-4}$.
	
	The temperature of the freeze-out can be easily obtained by considering in the framework of General Relativity the first Friedmann equation and the form of the radiation's energy density given in Eq. (\ref{enden}), 
	\begin{equation}
		H_{GR}=\sqrt{\frac{8\pi G}{3}\rho_r }= \left(\frac{\pi^2 g^*}{90}\right)^{1/2} \frac{T^2}{M_p}, \label{H}
	\end{equation}
	where $M_p=(8\pi G)^{-1/2}=1.22 \times 10^{19}\; \rm GeV$. Moreover, the time is related to the temperature by
	\begin{equation}
		\frac{1}{t} \simeq  \left(\frac{\pi^2 g^*}{90}\right)^{1/2} \frac{T^2}{M_p}. \label{param}
	\end{equation} 
	
	By taking into account that $H_{GR}=\Lambda(T_f)\approx qT_f^5$, the freeze-out temperature is given by
	\begin{equation}
		T_f=\left(\frac{\pi^2 g^*}{90 q^2 M_p^2}\right)^{1/6} \sim 0.0005\; \rm GeV.
	\end{equation}
	
	The abundance of $\rm ^4He$ is determined by analysing the primordial mass fraction estimate,
	\begin{equation}
		Y_p \equiv \lambda\; \frac{2 x(t_f)}{1+ x{t_f}},
	\end{equation}
	where $\lambda=e^{-(t_n-t_f)/\tau}$ represents the fraction of neutrons that decay into protons in a time interval $t \in [t_f,t_n]$, as $t_f$ is the freeze-out time and $t_n$, the time when the synthesis of helium nuclei occurs. The neutron mean life is given by $\tau = 870.2 \pm 15.8\; \rm s$ \cite{tau}. The function $x(t_f)=e^{-Q/T_f}$ resembles the neutron to proton equilibrium ratio.
	
	The freeze-out temperature varies and as a consequence, a quantity known as the deviation from the primordial mass fraction is derived,
	\begin{equation}
		\delta Y_p = Y_p \left[\left(1-\frac{Y_p}{2\lambda}\right) \ln \left(\frac{2\lambda}{Y_p}-1\right) - \frac{2t_f}{\tau}\right]\frac{\delta T_f}{T_f}.
	\end{equation}
	
	The variation of the temperature of nucleosynthesis is considered to be null $\delta T(t_n)=0$, as $T_n$ is fixed by the binding energy of deuterium \cite{Torres, Lambiase-2005}. The freeze-out temperature deviation is considered to have an upper limit given by
	\begin{equation}
		\left|{\frac{\delta T_f}{T_f} }\right| < 4.7 \times 10^{-4}. \label{Tf}
	\end{equation}
	
	\section{Cosmological models and Big Bang Nucleosynthesis in the presence of  a Weyl boundary}\label{numerical1}
	
	In this Section, we reformulate the basic evolution equations of the three cosmological  models proposed in Section~\ref{dyn} in a dimensionless form, which is more suitable for numerical computations. After reformulating the generalized Friedmann equations  as  three dimensionless systems of ordinary differential equations, from these equations we obtain the effective energy density and pressure of the primordial plasma in which the BBN processes took place. These physical quantities represent the basic input for the numerical comparison of the models with the BBN data. We also briefly review the first order approach to BBN, and the constraints on the energy density and parameter of the effective equation of state. In particular, the equation of state for our system, characterized by the $\rho_w$ and $p_w$ quantities, is capable of describing dark energy-like behaviour, which can naturally emerge from the early dynamics of the Universe.
	
	\subsection{Dimensionless form of the generalized Friedmann equations}
	
	In the following our main goal is to analyze the impact of the geometric terms on the evolution of the Universe with a Weylian boundary within the BBN paradigm. The specific cosmological framework governing the early Universe influences the abundances of light elements. In Section \ref{dyn}, we have proposed three such frameworks, and here, we evaluate the effects of radiation, scalar field and geometric effects on the nucleosynthesis process.

	We use the following set of dimensionless variables $(h, \theta, r_m,v(\phi),\lambda)$, defined as
	\begin{equation}
		H=H_0h, t = \frac{\tau}{H0},\rho _m=\frac{3H_0^2}{8\pi G}r_r, v(\phi)=\frac{V(\phi)}{H_0^2},\lambda =\frac{\Lambda}{ H_0^2}. \label{adim}
	\end{equation}
	Hence the three cosmological models can be written in a dimensionless form, as follows.
	
\paragraph{Model I - dimensionless form.} The equations that govern the first cosmological model take the following form
	\begin{equation}\label{t1}
		\frac{d^2\phi}{d\tau ^2}+3h\frac{d\phi}{d\tau}+v'(\phi)=-\beta \left(\frac{d\phi}{d\tau}\right)^2-\beta v(\phi),
	\end{equation}
	\begin{equation}\label{t2}
		\frac{d^2\psi}{d\tau ^2}+3h\frac{d\psi}{d\tau}+\left(\frac{d\psi}{d\tau}\right)^2=\lambda,
	\end{equation}
	\begin{eqnarray}\label{t3}
		2\frac{dh}{d\tau}+3h^2&=&-r_r-e^{\beta \phi}\left[\frac{1}{2}\left(\frac{d\phi}{d\tau}\right)^2-v(\phi)\right]\notag \\ &+&\frac{\alpha ^2}{2}\left[\lambda -\left(\frac{d\psi}{d\tau}\right)^2\right],
	\end{eqnarray}
	\begin{equation}\label{t4}
		\frac{dr_r}{d\tau}+4hr_r=\frac{\beta }{6}\left(\frac{d\phi}{d\tau}\right)^3 e^{\beta \phi}-\frac{\alpha ^2}{2}h\left(\frac{d\psi}{d\tau}\right)^2.
	\end{equation}

	These equations provide the time evolution of a Universe in which both the inflaton field and the Weyl vector influence the cosmological dynamics.
	
\paragraph{Model II - dimensionless form. } This scenario resembles the standard cosmological evolution, and it is given in a dimensionless form by the equations
	\begin{equation}\label{t5}
		\frac{d^2\phi}{d\tau ^2}+3h\frac{d\phi}{d\tau}+v'(\phi)=-\beta \left(\frac{d\phi}{d\tau}\right)^2- \beta v(\phi),
	\end{equation}
	\begin{equation}\label{t6}
		\frac{d}{d\tau}\left[\frac{d^2\psi}{d\tau ^2}+3h\frac{d\psi}{d\tau}+\left(\frac{d\psi}{d\tau}\right)^2\right]+3h\left(\frac{d\psi}{d\tau}\right)^2=0,
	\end{equation}
	\begin{eqnarray}\label{t7}
		2\frac{dh}{d\tau}+3h^2&=&-r_r-e^{\beta \phi}\left[\frac{1}{2}\left(\frac{d\phi}{d\tau}\right)^2-v(\phi)\right]\notag \\ &+&\frac{\alpha ^2}{2}\left(\frac{d^2\psi}{d\tau ^2}+3h\frac{d\psi}{d\tau}\right),
	\end{eqnarray}
	\begin{equation}\label{t8}
		\frac{dr_r}{d\tau}+4hr_r=\frac{\beta }{6}\left(\frac{d\phi}{d\tau}\right)^3 e^{\beta \phi}.
	\end{equation}

In this model particle creation takes place due to the decay of the scalar field $\phi$ only.
	
\paragraph{Model III - dimensionless form.} The evolution of a Universe in which the creation of particles  is possible due to the evolution of the Weyl vector field only is given by the following equations
	\begin{equation}\label{t9}
		\frac{d^2\phi}{d\tau ^2}+3h\frac{d\phi}{d\tau}+v'(\phi)=-\beta \left(\frac{d\phi}{d\tau}\right)^2-\beta v(\phi),
	\end{equation}
	\begin{eqnarray}\label{t10}
		\frac{\beta }{6}\left(\frac{d\phi}{d\tau}\right)^3 e^{\beta \phi}&=&\frac{\alpha^2}{2}\frac{d}{d\tau}\left(\frac{d^2\psi}{d\tau ^2}+3h\frac{d\psi}{d\tau}+\left(\frac{d\psi}{d\tau}\right)^2\right) \notag \\ &+& \frac{\alpha^2}{2}h\left(\frac{d\psi}{d\tau}\right)^2,
	\end{eqnarray}
	\begin{eqnarray}\label{t11}
		2\frac{dh}{d\tau}+3h^2&=&-r_r-e^{\beta \phi}\left[\frac{1}{2}\left(\frac{d\phi}{d\tau}\right)^2-v(\phi)\right]\notag \\&+&\frac{\alpha ^2}{2}\left(\frac{d^2\psi}{d\tau ^2}+3h\frac{d\psi}{d\tau}\right),
	\end{eqnarray}
	\begin{equation}\label{t12}
		\frac{dr_r}{d\tau}+4hr_r=\frac{\alpha^2}{3}\frac{d}{d\tau}\left(\frac{d^2\psi}{d\tau ^2}+3h\frac{d\psi}{d\tau}+\left(\frac{d\psi}{d\tau}\right)^2\right).
	\end{equation}
	
	These three distinct scenarios can be analyzed by considering different scalar field potentials $v(\phi)$. Consequently, we have chosen $V(\phi)$ as to have various forms, namely a null potential, 
\be
v(\phi)=0,  
\ee
a quadratic potential of the form 
\be
v(\phi)=m \phi ^2/2,
\ee
 a Higgs type potential 
 \be
 v(\phi)=\gamma \phi^2+\delta\phi^4, 
 \ee
 and finally an exponential potential, given by 
 \be
 v(\phi)=\sigma e^{-\mu \phi}.
 \ee

\subsection{BBN in cosmological models with Weylian boundary}

As previously mentioned, the early Universe primarily consisted of radiation, which satisfies the equation of state $\rho_r = p_r/3$. Therefore, the first Friedmann equation of the cosmological models with a Weylian boundary takes the general form
	\begin{equation}
		H^2 = \frac{8\pi G}{3}\left(\rho_r + \rho_w\right),
	\end{equation}
	where $\rho_w$ represents the contributions of the two scalar fields $\phi$ and $\psi$ to the first Friedmann equation~(\ref{F1}).
	
	We can assume that the Weyl effects are small as compared to the influence of the radiation sector, and as $H_{GR} \propto T^2$, we assume that in cosmological models with Weylian boundary the following approximation also holds, $H \propto T^2$, where $H$ is the Hubble function of the considered modified gravity models. Moreover, we follow an approach developed in \cite{Capozziello, fQ}. Within this framework, the Hubble function is expressed as
	\begin{eqnarray}
		H &=& H_{GR}\sqrt{1 + \frac{\rho_{w}}{\rho_r}} = H_{GR} + \delta H  \label{m2}\\
		\delta H &=& H_{GR}\left(\sqrt{1+\frac{\rho_{w}}{\rho_r}}-1\right). \label{m1}
	\end{eqnarray}
	
As the expression $\delta H$ is a deviation from the standard Hubble function $H_{GR}$, a deviation $\delta T_f$ will be induced as well from $T_f$, and since $H_{GR}=\Lambda(T_f)\approx qT_f^5$, 
	
	\begin{equation}
		\delta H = 5qT_f^4 \delta T_f,
	\end{equation}
	or
	\begin{equation}
		\frac{\delta H}{H} = 5 \frac{\delta T_f}{T_f}.
	\end{equation}
By taking into account Eq.~(\ref{H})  we obtain
	\begin{equation}
		\frac{\delta T_f}{T_f} \simeq \frac{\rho_{w}}{\rho_r}\frac{H_{GR}}{10q T_f^5} = \frac{\rho_w}{10q}\frac{1}{T_f^7} \left(\frac{10}{\pi^2 g^*}\right)^{1/2} \frac{1}{M_p}. \label{T}
	\end{equation}
	
	We can rewrite the previous relation by using Eq.~(\ref{enden}), so that the energy density of the geometric terms becomes
	\begin{equation}
		\rho_w = 10q T_f^7 M_p\sqrt{\frac{\pi^2 g^*}{10}} \frac{\delta T_f}{T_f}.
	\end{equation}
	 By making use of the upper bound provided by Eq.~(\ref{Tf}), the numerical values for $q = 9.6\times 10^{-10}$ GeV$^{-4}$, $g^*=10$ and $T_f = 0.0005$ GeV, we obtain the upper bound for the energy density of the geometric terms
	\begin{equation}
		 \rho_w < 1.35 \times 10^{-15}\;  \rm{GeV^4} = 1.35 \times 10^{-3} \; \rm{MeV^4}. \label{upper}
	\end{equation}
	
	This result imposes some strong constraints on any supplementary energy density considered present at the beginning of the Universe, and allows for small deviations from the standard expression of the radiation energy density.
	
	As the influence of the scalar field derived from the Weyl vector field $\psi$ is both relevant in the dynamical evolution of the Universe, the parameter $w_w$ of the equation of state for the geometric terms reads 
	\begin{equation}
		w_w = \frac{ p_w}{\rho_w } = \frac{p_\phi - \frac{\alpha^2}{2}\left(\ddot{\psi}+3H\dot\psi\right)}{\rho_\phi + \frac{\alpha^2}{2}\left(\ddot{\psi}+3H\dot\psi + \dot\psi^2\right)}, \label{w_w}
	\end{equation}
	which depends on how the scalar fields $\phi$ and $\psi$ contribute to the creation of matter and space. 

We adopt values of $w_w$ within the interval $ -1 < w_w < 1 $, which captures both dark energy-like $w_w \approx -1$ and radiation-like $w_w \approx 1/3$ behaviours for the geometric terms, accounting for the creation of particles, as well as for a large variety of dynamical behaviors. The imposed energy density in Eq.~(\ref{upper}), remains valid in all considered scenarios described in Section \ref{models}.

\section{Numerical implementation in PRyMordial library}\label{numerical}

In order to solve the three systems of equations describing the three models for the early Universe we propose an approach that implements a Genetic Algorithm for determining the initial values of the systems of the differential equations. By integrating the results into the python library \href{https://github.com/vallima/PRyMordial/tree/main}{\textcolor{blue}{PRyMordial}}, we can compare the predicted abundances of light elements against the standard Big Bang Nucleosynthesis predictions in an MCMC analysis. This comparison enables us to estimate the model parameters for different scalar potentials used in the modified gravitational framework. The methodology used for the numerical analysis is available in the \href{https://github.com/croi900/genesys}{\textcolor{blue}{genesys}} python program.
	
	In what follows, we will present a brief overview of the software package used in the calculation of the primordial abundances of light nuclei. This process was facilitated by the library \texttt{PRyMordial} \cite{primordial}. Essentially, the authors compute the abundances of light nuclei based on the thermonuclear rates provided in the NACRE II database \cite{nacre2}. 
	
	In \texttt{PRyMordial}, the nuclear processes are evaluated within a temperature range that starts at $T_\gamma = 10$ MeV and ends at keV scale. Initially, the neutrino species were in thermal equilibrium with the plasma, consisting of electrons coupled to photons, but as soon as the system cools at MeV scale and the nuclear freeze-out occurs, there exist two temperature regimes described by $T_\nu$ and $T_\gamma = T_e$, computed through systems of momentum-integrated Boltzmann equations. 
	The electroweak neutron-to-proton interconversions are calculated from nucleon abundances $Y_{n,p} = n_{n,p}/n_B$, which maintain chemical equilibrium before the decoupling of neutrinos. As the temperature falls below $T_f \simeq 0.5$ MeV, the $n \leftrightarrow p$  interconversion ceases and the thermodynamic background is integrated to obtain the scale factor of the universe $a$ as a function of the plasma temperature. Consequently, by the use of the Friedmann equations in the evaluation of the plasma dynamics, the number of baryons can be obtained and therefore, the light nuclei abundances can be estimated. 
	
The Standard Model energy density can be extended to include additional terms, which is facilitated by a feature in \texttt{PRyMordial} that allows us to introduce the contribution of the boundary terms into the thermodynamic background. The effect of $\rho_w$ is noticeable both on the temperature functions $T_\nu(T_\gamma)$ and the scale factor evolution, so that eventually the abundances of elements will slightly deviate from the standard values. 
	
	The additional energy density of the geometric terms can be analysed by solving the three systems of non-linear differential equations Eqs.~(\ref{t1}-\ref{t4}), (\ref{t5}-\ref{t8}) and (\ref{t9}-\ref{t12}) and by  imposing the nuclear freeze-out constraint given in Eq.~(\ref{upper}) and the equation of state given in Eq.~(\ref{w_w}). 
	The time dependent results can be parameterized as functions of temperature $T \in [10, 10^{-4}]\; \rm MeV$, in accordance with the decreasing temperature scale defined in \texttt{PRyMordial}. 
	
	However, as the three models define initial value boundary problems, we cannot solve the systems of non-linear differential equations without knowing the initial values of the inflationary functions, namely $\phi(0) = \phi_0$, $\dot{\phi}(0) = \phi_{01}$,  $\psi(0) = \psi_0$, $\dot{\psi}(0) = \psi_{01}$, $\ddot{\psi}(0) = \psi_{02}$, and the constant associated with the Weyl vector field $\alpha$, respectively. The dissipation constant $\beta$ 
	is obtained as a constraint from the first Friedmann equation in Eq.~(\ref{F1}),  having the following expression,
	\begin{equation}
		\beta = \frac{1}{\phi(0)}\ln\left(\frac{6 - \alpha^2(\psi^{\prime\prime}(0) + 3h(0)\psi^\prime(0) + (\psi^\prime(0))^2)}{2\left[\frac{\phi'(0)^2}{2} + v(\phi(0))\right]}\right)
	\end{equation}
	
	and the $\lambda$ parameter is obtained for the particular case in which the Weyl scalar field is conserved.

	Given the complexity of this system, the total number of unknown parameters is large, namely six initial conditions, two potential parameters, and the coupling $\alpha$. In contrast, the number of available observational constraint is limited to three (hydrogen, deuterium, helium-3 and helium-4, excluding lithium-7 as it is highly uncertain and cannot be used in estimating model parameters). This leads to a negative number of degrees of freedom, making any kind of analysis invalid.
	
	To address this issue and reduce the dimensionality of the parameter space, we use a continuous Genetic Algorithm, which is described in detail in the following section. This computational framework allows us to efficiently search for suitable combinations of initial conditions, reducing the number of free parameters to one or two, depending on the scalar potential model used.

\subsection{Initial parameter estimation with Genetic Algorithms} \label{genetic_alg}
		
	Genetic Algorithms are optimization routines, which search for the best output or minimal cost from all possible values without the need of taking derivatives of the cost function. Unlike traditional downhill methods which often converge in local minima (e.g. gradient descend, Nelder-Mead Simplex Method), Genetic Algorithms focus on generating new points in the direction of optimal variables by mimicking natural methods found in evolutionary processes in order to find a global minima. Genetic Algorithms were extensively studied, but only some remarkable early works are briefly mentioned here \cite{Holland, Goldberg, GA1, GA2, GA3}.
	
	Natural optimization methods require an organism, represented by the results of iterations, a cost function which resembles the survival of the organism, achieved by minimizing the genetical traits that disfavour a maximal fitness function, i.e. a maximum adaptation to the environment, defined by constraints for the initial parameters \cite{genetic_alg}. Genetic Algorithms rely on finding the best genes, which in our case are the initial values for the system of differential equations. A sequence of genes forms a chromosome, the parameter set $\{ \phi_0, \phi_{01}, \psi_0, \psi_{01}, \psi_{02} / \lambda, r_{r0}, h_0, \alpha\}$ which gives valid solutions to Eq.~(\ref{t1}-\ref{t12}), with the fifth parameter being either $\psi_{02}$ or $\lambda$ depending on the model. Similarly to genetic processes, new chromosomes are generated from parent chromosomes that are best adapted to environmental conditions. The parameters which minimize the cost function serve as gene sources for the next generation through a cross-over function. The evolutionary process consists of a mutation mechanism that alter genes randomly in order to escape from local minima. This process along with the gene flow provided by mating and natural selection given by fitness function characterize the dynamic aspect of the algorithm.

	\begin{table*}[htbp!]
		\centering
		\begin{tabular}{|c| c| c| c| c| c| c| c| c| c| c|}
			\hline\hline
			Model & $\phi_0$ & $\phi_{01}$ & $\psi_0$ & $\psi_{01}$ & $\alpha$ & $\lambda/ \psi_{02}$ & Method & rtol & atol & Range \\
			\hline
			 I & $-0.0261$ & $-0.3796$ & $-0.2597$ & $0.0738$ & $-0.1899$ & $0.5648$ & BDF & $10^{-3}$ & $10^{-6}$ & $[-1, 1]$ \\
\hline
			 II & $-0.0369$ & $-0.0174$ & $0.0495$ & $-0.0475$ & $0.0239$ & $0.0373$ & BDF & $10^{-3}$ & $10^{-6}$ & $[-0.1, 0.1]$ \\
\hline
			 III & $-0.0094$ & $-0.0009$ & $-0.0880$ & $0.0155$ & $0.0233$ & $-0.0495$ & BDF & $10^{-2}$ & $10^{-5}$ & $[-0.1, 0.1]$ \\
			\hline
		\end{tabular}
		\caption{Initial parameters found with the Genetic Algorithm for the three warm inflationary models, including solver settings and parameter space bounds. For the first model,  $\lambda$ is the sixth parameter, whereas in the second and third models, the parameter is $\psi_{02}$, respectively.}
		\label{model_parameters}
	\end{table*}

	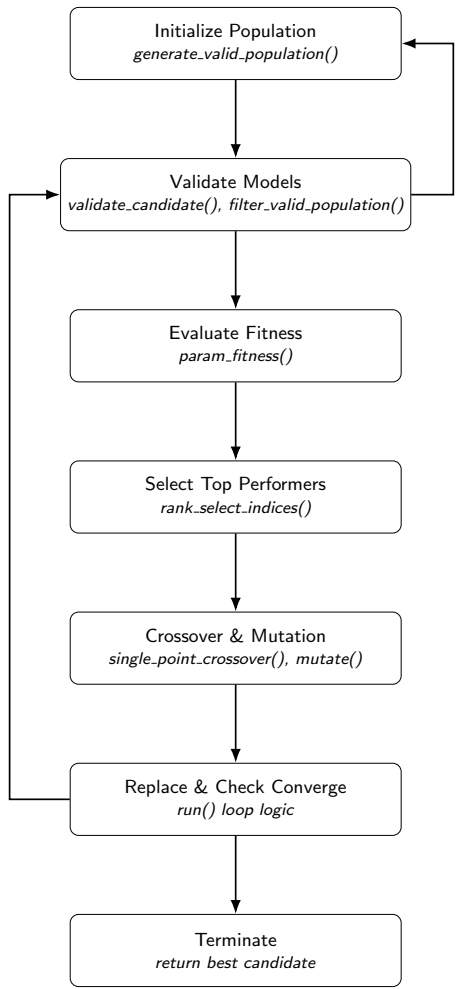
\begin{figure}[htbp!]
		\centering
		\scalebox{0.8}{
		\begin{tikzpicture}[
			node distance=1.3cm,
			box/.style={rectangle, draw, rounded corners, align=center, minimum width=5.5cm, minimum height=1.2cm, font=\sffamily},
			arrow/.style={-Latex, thick}
			]
			
			\node[box] (init) {Initialize Population\\\footnotesize\textit{generate\_valid\_population()}};
			\node[box, below=of init] (validate) {Validate Models\\\footnotesize\textit{validate\_candidate(), filter\_valid\_population()}};
			\node[box, below=of validate] (fitness) {Evaluate Fitness\\\footnotesize\textit{param\_fitness()}};
			\node[box, below=of fitness] (select) {Select Top Performers\\\footnotesize\textit{rank\_select\_indices()}};
			\node[box, below=of select] (crossover) {Crossover \& Mutation\\\footnotesize\textit{single\_point\_crossover(), mutate()}};
			\node[box, below=of crossover] (replace) {Replace \& Check Converge\\\footnotesize\textit{run() loop logic}};
			\node[box, below=of replace] (terminate) {Terminate\\\footnotesize\textit{return best candidate}};
			
			\draw[arrow] (init) -- (validate);
			\draw[arrow] (validate) -- (fitness);
			\draw[arrow] (fitness) -- (select);
			\draw[arrow] (select) -- (crossover);
			\draw[arrow] (crossover) -- (replace);
			\draw[arrow] (replace) -- (terminate);
			
			\draw[arrow] (replace.west) |- ++(-1,0) |- (validate.west);
			
			\draw[arrow] (validate.east) |- ++(0.7,0) |- (init.east);
			
		\end{tikzpicture}
	}
		\caption{Flow diagram of the Genetic Algorithm routine.}
		\label{GA}
	\end{figure}

	We implemented a Continuous Genetic Algorithm available in \texttt{genesys}, whose workflow is presented in Fig.~(\ref{GA}). The initial parameter values are selected within a given interval \{\texttt{varlo}, \texttt{varhi}\} and are continuously regenerated within \texttt{generate\_valid\_population} function until initial sets of parameters are found. The parameters are validated if they solve all four types of scalar potentials within a model, this being handled in \texttt{validate\_candidate} and \texttt{filter\_valid\_population} functions. Afterwards, the valid parameters are evaluated in \texttt{param\_fitness} and ranked with respect to variations in $w_w$ and $\rho_w(t_f)$ at freeze-out temperature or non-physical negative values for the functions. 
	New generations of parameters are obtained within the crossover operation in \texttt{single\_point\_crossover} function, encapsulating the following strategy: for each pair of selected parents, a random point along their parameter sequence is chosen, and their components are swapped beyond this point. In addition to crossover, a mutation function \texttt{mutate} is applied to maintain diversity in the population and escape local minima. 
	The results of the Genetic Algorithm are presented in Table \ref{model_parameters}.

	The best-fit initial values for the functions are initialized in a general Metropolis-Hastings MCMC algorithm, which is implemented to establish the parameter space of the $v(\phi)$ potentials, specifically $\{ m, \gamma, \delta, \sigma, \mu\}$ values. These parameters are estimated using the \texttt{PRyMordial} library by comparing the predicted light nuclei abundances, which are generated for a plasma containing contributions coming from the modified gravity framework. The newly generated abundances are evaluated against the SBBN predictions, which serve as observational data for the analysis, as it will be presented in the following section.

	\subsection{ Parameter Constraints from BBN limit} \label{alphapy}

	We conduct a further analysis to determine the limits imposed by the BBN constraints on the model parameters. Specifically, we seek to obtain a physical limit for fundamental quantities, such as the  Weyl coupling parameter, which can be inferred from Eq.~\ref{upper}. We therefore constructed a second Genetic Algorithm that treats the optimized parameter values from Table~\ref{model_parameters} as mean values of normal distributions and searches for optimal standard deviations that maximize the probability of obtaining physically valid parameter combinations from a BBN standpoint.
	
	The \texttt{alpha.py} file from the GitHub repository implements this genetic algorithm for determining the distributions of the null-potential models' initial parameters. Starting from the optimized values given by the primary GA, we sample randomly from normal distributions centred on these parameter values. The algorithm initializes a population of candidate $\sigma$ values, which represents standard deviations for the parameter distributions, computes fitness using a Monte Carlo approach, selects the best solutions as parents, creates offspring through crossover and mutation, replaces the previous population with the new generation, and repeats until convergence.
	
	Each genome represents a vector of standard deviation values corresponding to each model parameter. The genomes are ranked by a Monte Carlo process where a number of $T$ samples are drawn from each parameter's distribution. These are further validated using nearly the same technique as the initial parameter determination GA, with the fitness function defined as
	\begin{equation}
		f(\sigma)= \frac{F}{T} \times \frac{\sigma}{L},
	\end{equation}
	where $F$ represents the number of valid samples, $T=100$ is the total number of MC samples, and $L$ is the lower bound of allowed values for $\sigma$. This approach searches for optimized standard deviations for each parameter, with a fitness function defined as the maximized ratio $F/T$ of obtaining valid values across Monte Carlo samples. The term $\sigma/L$ maximizes distribution width, preventing results where the distribution becomes indistinguishable from sampling the mean indefinitely. This term contributes weakly when $\sigma$ approaches the lowest accepted limit, and we set $L= 0.001$ for the first two models and $L=0.00001$ for the third model, as it presented smaller parameter values. We thus search for the largest parameter intervals consistent with BBN limits while maintaining physically meaningful results.
	
	Due to the large number of system parameters and the fact that the genome consists of standard deviations, the Genetic Algorithm has to account for interdependencies between the parameter distributions, introducing large amounts of complexity to the problem. Moreover, the process involves drawing $T=100$ MC samples for six parameters and evaluate the fitness function accordingly, making the analysis computationally expensive. As a consequence, the introduction of random $\sigma$ values implied in a mutation process could make the system diverge fast, compromising the analysis. Therefore, we introduced a heuristic mutation strategy that varies each parameter's standard deviation within 10\%, maintaining a stable convergence path. The crossover is implemented in the same manner as for the first GA, described in Section \ref{genetic_alg}.
	
	The results of this Genetic Algorithm are as follows, with parameter estimates presented as 95\% confidence intervals:
	
	\noindent Model I:
	\begin{align*}
		-0.0532 &< \phi_0 < 0.0010 \\
		-0.7955 &< \phi_{01} < 0.0363 \\
		-2.2197 &< \psi_0 < 1.7003 \\
		-1.8862 &< \psi_{01} < 2.0338 \\
		-0.2908 &< \alpha < -0.0890 \\
		0.1116 &< \lambda < 1.0180
	\end{align*}
	Model II:
	\begin{align*}
		-1.8524 &< \phi_0 < 1.7786 \\
		-1.0245 &< \phi_{01} < 0.9897 \\
		-1.6585 &< \psi_0 < 1.7575 \\
		-0.1278 &< \psi_{01} < 0.0328 \\
		-0.0545 &< \alpha < 0.1023 \\
		-0.1505 &< \psi_{02} < 0.2251
	\end{align*}
	Model III:
	\begin{align*}
		-0.2027 &< \phi_0 < 0.1839 \\
		-0.0012 &< \phi_{01} < -0.0006 \\
		-0.2380 &< \psi_0 < 0.0620 \\
		-0.0175 &< \psi_{01} < 0.0485 \\
		-0.0014 &< \alpha < 0.0480 \\
		-0.1525 &< \psi_{02} < 0.0535
	\end{align*}
	
	For some intervals, the standard deviations are considerably larger than the parameter means found in Table~\ref{model_parameters}, indicating that the respective parameter can vary significantly in the ODE solving process without compromising the physicality of the results. Moreover, the intervals for Model III are tighter as the interval for allowed standard deviations in the Genetic Algorithm is smaller.
	
	The Weyl coupling parameter $\alpha$ has different values across the three Warm Weyl models. The first model exhibits a strong negative coupling parameter, $\alpha_{I}=-0.1899$, indicating that the Weyl vector dynamics opposes radiation production, while the other two models show weak positive values, $\alpha_{II} = 0.0239$ and $\alpha_{III}=0.0233$, respectively . The analysis reveals that Model I allows the broadest coupling variations (from -0.29 to 0.09) for maintaining stable ODE solutions.
	Models III requires tighter constraints (from -0.0014 to 0.0480), showing that the Weyl effects need must remain within tighter bounds to sustain radiation production.

	\section{Bayesian estimation of the scalar field potentials in Weyl boundary gravity}\label{numerical2} 
	
	A Markov Chain Monte Carlo approach consists of a class of methods that are used in drawing samples from distributions whose properties are not directly known. This method is inherently Bayesian and it is advantageous as it involves randomly selecting samples from probability distributions based on the original data, and then evaluating the models derived from these samples in comparison to the actual data.
		
	The probability that the parameters properly describe the data based on the model is $P(\theta|D)$ and is calculated by using Bayes' theorem, 
	\begin{equation}
			P(\theta|D) = \frac{P(D|\theta) P(\theta)}{P(D)},
	\end{equation}
	where $P(D|\theta)$ is the likelihood,  $P(\theta)$ is called prior and $P(D)$ is the evidence and $P(\theta|D)$ is called the posterior, which the MCMC procedure tries to estimate \cite{mc, MCMC}.

	\subsection{Metropolis-Hastings parameter estimation}
	
	In the MCMC parameter fitting process, the estimation of scalar potential values is obtained by choosing a $\chi^2$ function which quantifies the discrepancy between the model and observed data. The $\chi^2$ statistic was chosen to have the following form,
	\begin{equation}
		 \chi^2 = \sum_{i} \frac{(X_{i,obs} - X_i)^2}{\sigma_i^2}, \label{chi2}
	\end{equation}
	where $X_{i, obs}$ are the mean observed values of the abundances and $\sigma_i^2$ are the uncertainties in the observed values, taken from \cite{pdg}, respectively. $X_i$ are the predicted values of the nuclear abundances. Their values are displayed in Table \ref{abundances}. The Lithium abundance is not considered in the $\chi^2$ function as it is highly fluctuating within the \texttt{PRyMordial} framework and it would be impractical to use in our calculations of scalar potential coefficients.
	
	\begin{table}[h!]
		\centering
		\begin{tabular}{|c| c| c|}
			\toprule
			Abundance & $X_{i, obs}$ & $\sigma_i^2$ \\
			\midrule
\hline
			$\rm Y_p$ & 0.245 & 0.003 \\
\hline
			$\rm D/H \times 10^5$ & 2.547 & 0.029 \\
\hline
			$\rm ^3He/H \times 10^5$ & 1.08 & 0.12 \\
			\bottomrule
\hline
		\end{tabular}
		\caption{Primordial abundance parameters with their observed average values and standard deviations.}
		\label{abundances}
	\end{table}

	\begin{table*}[htbp!]
		\centering
		\renewcommand{\arraystretch}{1.3}
		\setlength{\tabcolsep}{4pt}
		\begin{tabular}{|l|c|c|c|c|c|c|}
			\hline
			Model &
			$\chi^2_\nu$ &
			AIC &
			BIC &
			$Y_p$ &
			D/H ($\times10^5$) &
			$^3$He/H ($\times10^5$) \\
			\hline
			\texttt{M1VPHI2}  & 0.6816 & 3.3632 & 2.4618 & $0.246106^{+0.000405}_{-0.000405}$ & $2.3231^{+0.6213}_{-0.6213}$ & $1.0399^{+0.0757}_{-0.0757}$ \\
\hline
			\texttt{M1VPHI24} & 1.3582 & 5.3582 & 3.5554 & $0.246427^{+0.003187}_{-0.003187}$ & $2.3484^{+0.5353}_{-0.5353}$ & $1.0585^{+0.0645}_{-0.0645}$ \\
\hline
			\texttt{M1VEXP}   & 1.4118 & 5.4118 & 3.6090 & $0.246998^{+0.000027}_{-0.000027}$ & $2.5309^{+0.1005}_{-0.1005}$ & $1.0465^{+0.0002}_{-0.0002}$ \\
\hline
			\texttt{M2VPHI2}  & 0.3384 & 2.6768 & 1.7726 & $0.246998^{+0.000003}_{-0.000003}$ & $2.5302^{+0.0144}_{-0.0144}$ & $1.0464^{+0.0002}_{-0.0002}$ \\
\hline
			\texttt{M2VPHI24} & 0.6802 & 4.6802 & 2.8777 & $0.246998^{+0.000001}_{-0.000001}$ & $2.5303^{+0.0139}_{-0.0139}$ & $1.0464^{+0.0002}_{-0.0002}$ \\
\hline
			\texttt{M2VEXP}   & 0.6920 & 2.6920 & 2.8393 & $0.246986^{+0.002792}_{-0.002792}$ & $2.5281^{+0.0589}_{-0.0589}$ & $1.0464^{+0.0210}_{-0.0210}$ \\
\hline
			\texttt{M3VPHI2}  & 0.3714 & 2.7429 & 1.8415 & $0.246997^{+0.000000}_{-0.000000}$ & $2.5318^{+0.0002}_{-0.0002}$ & $1.0465^{+0.0001}_{-0.0001}$ \\
\hline
			\texttt{M3VPHI24} & 0.7717 & 4.7717 & 2.9683 & $0.246996^{+0.000000}_{-0.000000}$ & $2.5314^{+0.0012}_{-0.0012}$ & $1.0464^{+0.0001}_{-0.0001}$ \\
\hline
			\texttt{M3VEXP}   & 0.6971 & 2.8974 & 2.8943 & $0.246998^{+0.000000}_{-0.000000}$ & $2.5390^{+0.0001}_{-0.0001}$ & $1.0465^{+0.0002}_{-0.0002}$ \\
			\hline
		\end{tabular}
		\caption{Summary of model statistics for different scalar potentials in the cosmological evolution in the presence of Weyl boundary terms,  and the primordial abundance predictions (excluding $^7$Li/H).}
		\label{results}
	\end{table*}

	The MCMC procedure starts by imposing the initial values for the differential systems of equations obtained through a Genetic Algorithm described in the previous subsection. Following, the non-zero scalar potential types are considered for which the additional contributions to the plasma are calculated as $\rho_w(t),\; p_w(t),\; \text{and}\; d\rho_w(t)/dt$, computed by analysing the behaviour of the warm inflationary functions. Subsequently, these quantities are parameterized as functions of temperature, because \texttt{PRyMordial} imposes a temperature dependency for all new physics contributions. 
	
\begin{figure*}[htbp!] 
		\centering
		\includegraphics[scale = 0.7]{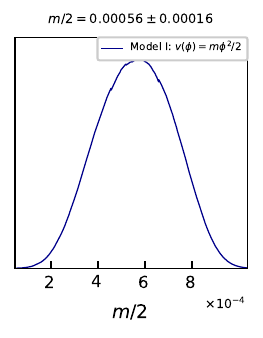}
		\includegraphics[scale = 0.7]{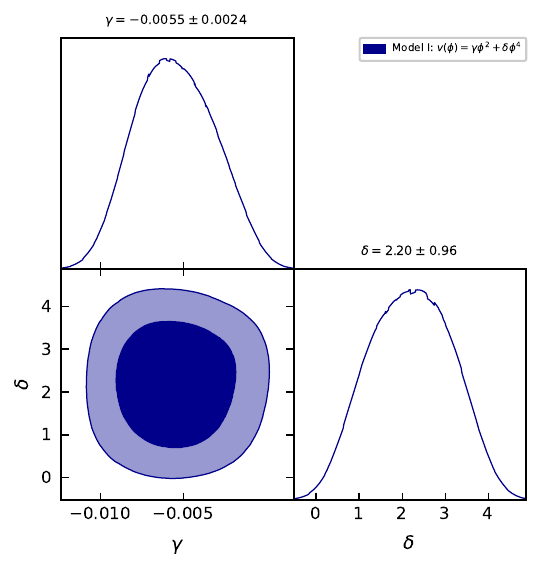}
		\includegraphics[scale = 0.7]{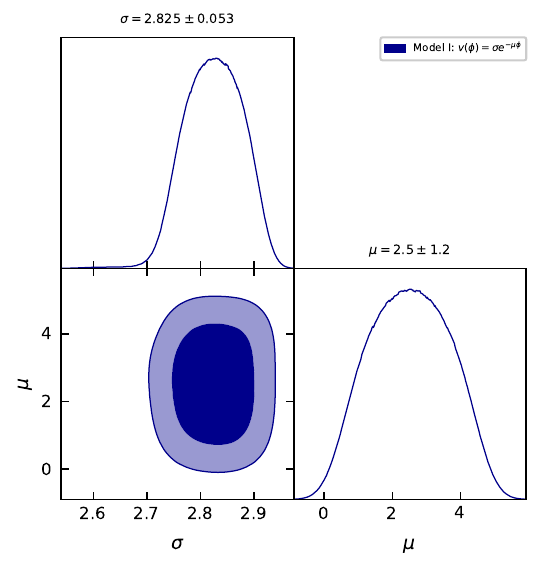}
		\includegraphics[scale = 0.7]{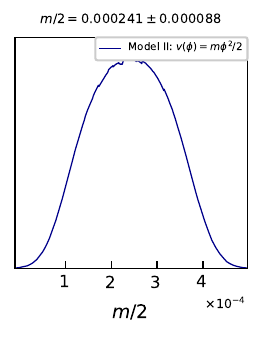}
		\includegraphics[scale = 0.7]{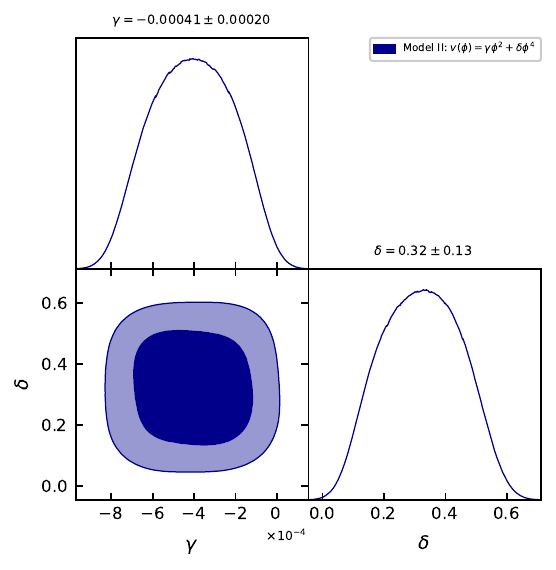}
		\includegraphics[scale = 0.7]{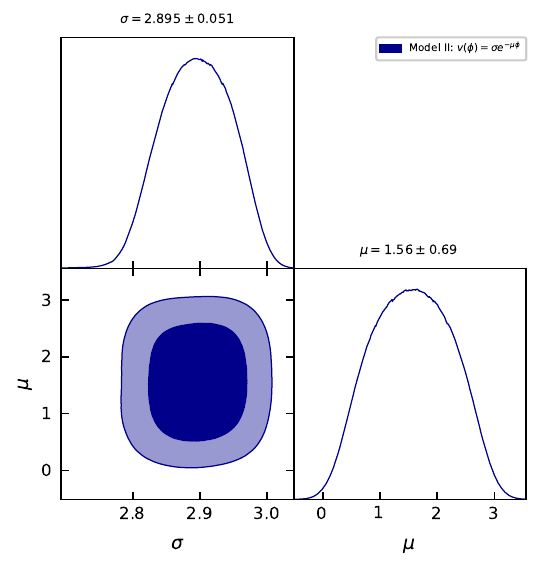}
		\includegraphics[scale = 0.7]{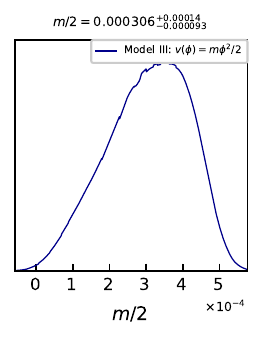}
		\includegraphics[scale = 0.7]{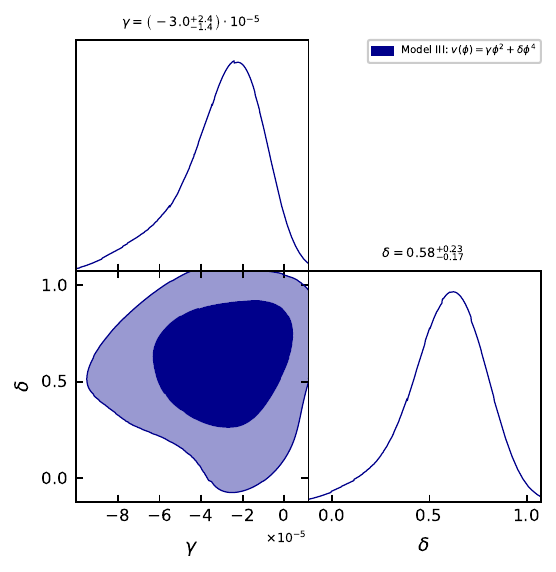}
		\includegraphics[scale = 0.7]{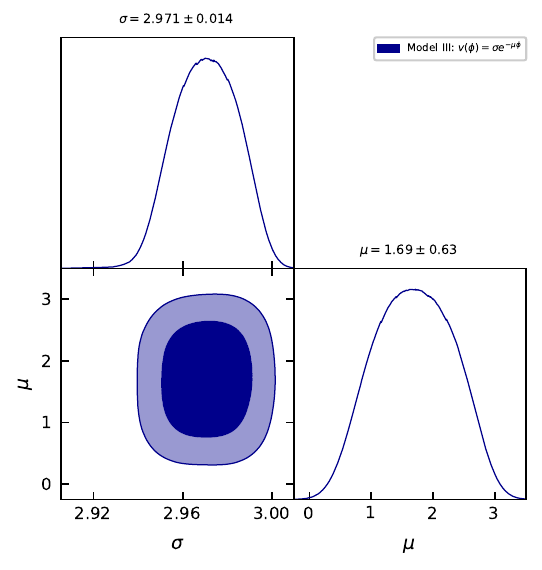}
		\caption{The posterior distributions of scalar potential parameters for the warm inflationary frameworks with boundary contribution, estimated with BBN constraints from primordial abundances.}
		\label{results_potentials}
	\end{figure*}

	 We employed a Metropolis-Hastings sampling method that accepts or rejects parameters by computing the value of the $\chi^2$ function, given in Eq.~\ref{chi2}.  
	 The simulation was implemented by using the \texttt{emcee} python library and the results of the nuclei abundances are displayed in Table \ref{results}.  The estimated posterior distributions for the three types of scalar potentials are presented in Fig.~\ref{results_potentials}. The corner plots are generated using the Stein thinning procedure described in \cite{thinning}, with thinning parameters obtained from autocorrelation time, after applying a 10\% discard.

	 \begin{table}[htbp!]
	 	\centering
	 	
	 	\renewcommand{\arraystretch}{1.1}
	 	\begin{tabular}{|l|c|c|c|c|c|}
	 		\hline
	 		\textbf{Model} 
	 		& $\mu_{N_{\text{eff}}}$ 
	 		& $\sigma_{N_{\text{eff}}}$ 
	 		& $\Delta N_{\text{eff}}$ 
	 		& $\mu_{\Omega_\nu h^2}$ 
	 		& $\sigma_{\Omega_\nu h^2}$ \\
	 		\cline{5-6}
	 		& & & & \multicolumn{2}{c|}{$\times 10^6$} \\
	 		\hline
	 		\texttt{M1VPHI2}   & 3.044394 & 0.000001 & 0.000394 & 5.699409 & 0.000001 \\
	 		\texttt{M1VPHI24}  & 3.044368 & 0.000075 & 0.000368 & 5.699357 & 0.000145 \\
	 		\texttt{M1VEXP}    & 3.045109 & 0.000583 & 0.001109 & 5.699868 & 0.000373 \\
	 		\texttt{M2VPHI2}   & 3.044411 & 0.000004 & 0.000411 & 5.699421 & 0.000002 \\
	 		\texttt{M2VPHI24}  & 3.044402 & 0.000005 & 0.000402 & 5.699416 & 0.000003 \\
	 		\texttt{M2VEXP}    & 3.045079 & 0.000668 & 0.001079 & 5.699849 & 0.000427 \\
	 		\texttt{M3VPHI2}   & 3.044520 & 0.000016 & 0.000520 & 5.699491 & 0.000010 \\
	 		\texttt{M3VPHI24}  & 3.044436 & 0.000017 & 0.000436 & 5.699438 & 0.000011 \\
	 		\texttt{M3VEXP}    & 3.045023 & 0.000605 & 0.001023 & 5.699813 & 0.000387 \\
	 		\hline
	 	\end{tabular}
	 	\caption{ Summary statistics for $N_{\text{eff}}$ and $\Omega_\nu h^2$ across all Weyl models. The values of $\Omega_\nu h^2$ are shown in units of $10^{-6}$.}
	 	\label{neff}
	 \end{table}
	 
	 Besides the estimation of nuclear abundances, the effective number of neutrino species $N_{\text{eff}}$
	 is also computed along with the neutrino density parameter $\Omega_\nu h^2$, displayed in Table~\ref{neff}. These results are obtained from a Monte Carlo analysis where both the scalar potential parameters and PRyMordial's initial parameters (nuclear reaction rates, neutron mean lifetime and baryon-to-photon ratio) were sampled from their posterior distributions. It can be observed that their values are in agreement with most recent results in \cite{yeh}, $\Delta N_{\text{eff}} < 0.18$ at 2$\sigma$ significance level. Our results show noticeable but minimal deviation from the standard value $N_{\text{eff}}^{SM}$ = 3.044.

	 The predictions of our models are consistent with the cosmological constraints derived from Planck+BAO measurements, which find $N_{\text{eff}} = 2.99 \pm 0.17$ at 68\% CL \cite{bao}. Similarly, the helium-4 abundance $Y_p\approx 0.247$ and deuterium ratio $D/H \approx 2.53 \times 10^{-5}$ resulted from our analysis agree with the value inferred when Planck CMB data are combined with BAO and a BBN prior, $Y_p \approx 0.2471 \pm 0.0002$ at 95\% CL \cite{bao}, while our deuterium ratio matches the BBN-calibrated value obtained at the Planck baryon density $D/H \approx (2.527 \pm 0.030) \times 10^{-5}$ at 68\% confidence level \cite{cooke}. These comparisons show that the early-time energy density modification in our framework remains fully compatible with late-time CMB and BAO datasets.

	 \subsection{Convergence Analysis}
	 
	  We used the Gelman-Rubin $\hat{R}$ statistic to provide a diagnostic convergence for the MCMC chains, which evaluates the ratio of the variance within the sample chain and the variance within each chain \cite{gelman}. The R statistic is computed as 
	 \begin{equation}
	 	\hat{R} = \sqrt{\frac{\hat{V}}{\hat{W}}},
	 \end{equation}
	 where  $\hat{V} = \frac{n-1}{n}W + \frac{m+1}{mn}B$ is the posterior variance estimate, $\hat{W} = \frac{1}{m}\sum_{j=1}^{m} s_j^2$ is the variance within the chain, and $s_j^2$ is the variance of chain $j$. The variance within each chain is given by $B = \frac{n}{m-1}\sum_{j=1}^{m}(\bar{\theta}j - \bar{\theta})^2$, where $\bar{\theta}_j$ is the mean of chain $j$ and $\bar{\theta}$ is the overall mean, $m$ is the number of chains, and $n$ is the number of samples per chain. 
	 
	 We ran $m = 1$ chain for the quadratic potential models and $m = 2$ chains for the Higgs-type and exponential models, representing the number of parameters constrained in each case. By using a total number of $n = 3064$ steps per chain, we obtained the following results for the $\hat{R}$ test, listed in Table~\ref{Rtest}.
	 
	 \begin{table}[htbp!]
	 	\centering
	 	
	 	\renewcommand{\arraystretch}{1.4}
	 	\begin{tabular}{|l|c|c|}
	 		\hline
	 		\textbf{Model} & \textbf{Param 1} & \textbf{Param 2} \\ \hline
	 		\texttt{M1VPHI2} & $R(m) = 1.014$ & N/A \\
	 		$v(\phi) = \frac{1}{2}m\phi^2$ & & \\ \hline
	 		\texttt{M1VPHI24} & $R(\gamma) = 1.001$ & $R(\delta) = 1.002$ \\
	 		$v(\phi) = \gamma \phi^2 + \delta \phi^4$ & & \\ \hline
	 		\texttt{M1VEXP} & $R(\sigma) = 1.000$ & $R(\mu) = 1.000$ \\
	 		$v(\phi) = \sigma e^{-\mu \phi}$ & & \\ \hline
	 		\texttt{M2VPHI2} & $R(m) = 1.004$ & N/A \\
	 		$v(\phi) = \frac{1}{2}m\phi^2$ & & \\ \hline
	 		\texttt{M2VPHI24} & $R(\gamma) = 1.002$ & $R(\delta) = 1.002$ \\
	 		$v(\phi) = \gamma \phi^2 + \delta \phi^4$ & & \\ \hline
	 		\texttt{M2VEXP} & $R(\sigma) = 1.002$ & $R(\mu) = 1.000$ \\
	 		$v(\phi) = \sigma e^{-\mu \phi}$ & & \\ \hline
	 		\texttt{M3VPHI2} & $R(m) = 1.017$ & N/A \\
	 		$v(\phi) = \frac{1}{2}m\phi^2$ & & \\ \hline
	 		\texttt{M3VPHI24} & $R(\gamma) = 1.305$ & $R(\delta) = 1.799$ \\
	 		$v(\phi) = \gamma \phi^2 + \delta \phi^4$ & & \\ \hline
	 		\texttt{M3VEXP} & $R(\sigma) = 1.000$ & $R(\mu) = 1.006$ \\
	 		$v(\phi) = \sigma e^{-\mu \phi}$ & & \\ \hline
	 	\end{tabular}
	 	\caption{ Gelman–Rubin $\hat{R}$ diagnostic values for each model parameter.}
	 	\label{Rtest}
	 \end{table}

\begin{figure}
	 	\centering
	 	\includegraphics[scale = 0.348]{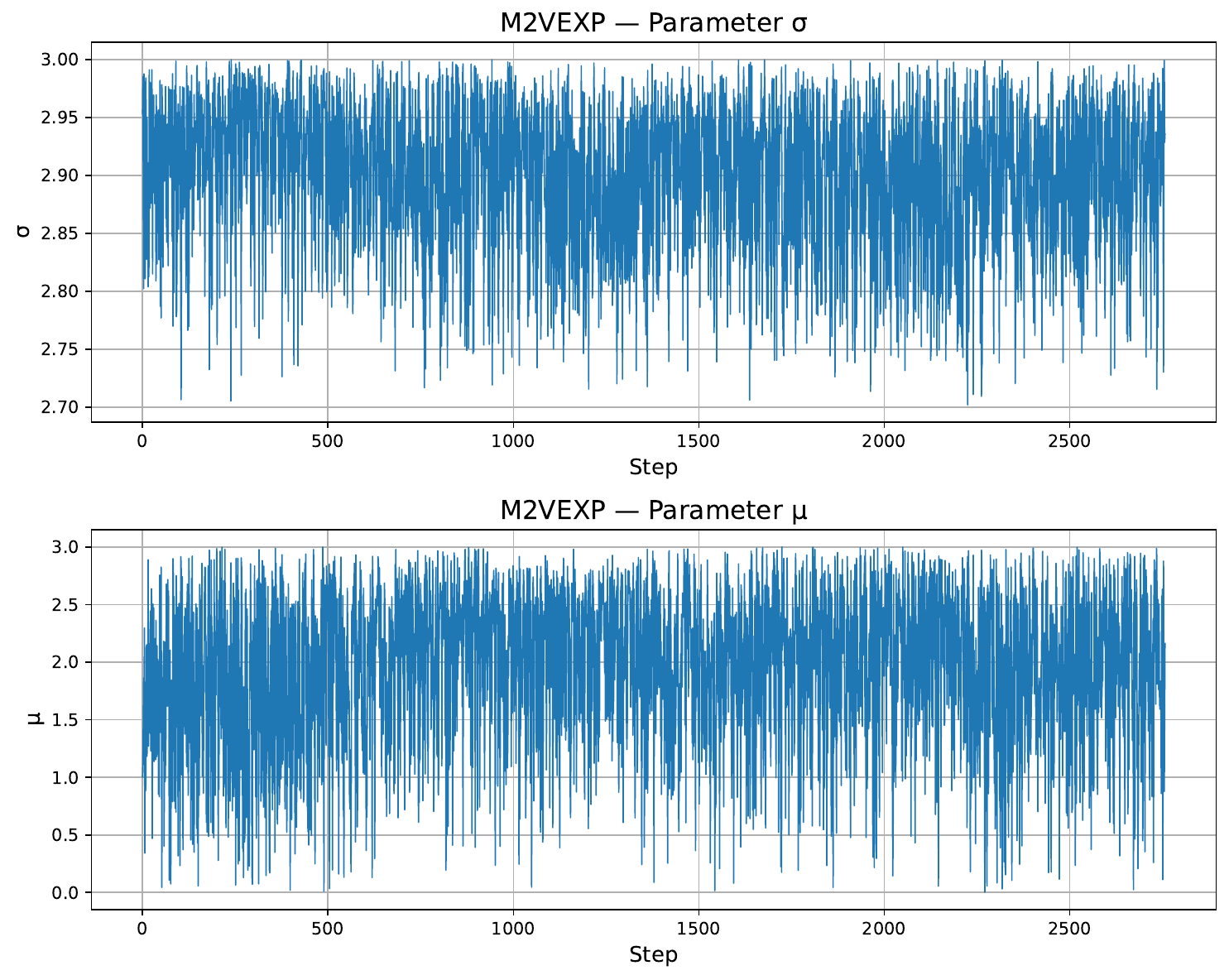}
	 	\caption{ Trace plots showing the evolution of the chains with increasing samples for model \texttt{M2VEXP}.}
	 	\label{trace}
	 \end{figure}

	 The Gelman-Rubin $\hat{R}$ values indicate that the chains converged successfully, as $\hat{R} \lesssim 1.02$, for all parameters except $\gamma$ and $\delta$ in \texttt{M3VPHI24} model, which show substantially larger values, $\hat{R} = 1.305$ and $\hat{R} = 1.799$. These outliers indicate a poor convergence for these parameters, as shown in the skewed posterior distributions present in Fig.~\ref{corners}. The convergence of the parameter chains can be also confirmed by analysing their trace plots, which show the evolution of the chains as sampling progresses. One representative trace plot is shown in Fig.~\ref{trace}, which illustrates that for model \texttt{M2VEXP} the chains reach stationarity for the $\sigma$ and $\mu$ parameters.
	 
	 The priors in our MCMC analysis are specified as uniform distributions with the bounds given in Table~\ref{priors}.  These intervals were specifically selected to avoid numerical instabilities in \texttt{PRyMordial} that result from the stiffness of the evaluated system and to avoid unphysical results.
	 
	 \begin{table}[h!]
	 	\centering
	 	
	 	\renewcommand{\arraystretch}{1.2}
	 	\begin{tabular}{|l|c|}
	 		\hline
	 		Model & Prior interval \\ \hline
	 		\texttt{M1VPHI2}   & $[-1,\,1]$ \\
	 		\texttt{M1VPHI24}  & $[-3,\,3]$ \\
	 		\texttt{M1VEXP}    & $[-5,\,5]$ \\
	 		\texttt{M2VPHI2}   & $[-0.5,\,0.5]$ \\
	 		\texttt{M2VPHI24}  & $[-4,\,4]$ \\
	 		\texttt{M2VEXP}    & $[-5,\,5]$ \\
	 		\texttt{M3VPHI2}   & $[-0.5,\,0.5]$ \\
	 		\texttt{M3VPHI24}  & $[-4,\,4]$ \\
	 		\texttt{M3VEXP}    & $[-5,\,5]$ \\
	 		\hline
	 	\end{tabular}
	 	\caption{ Prior bounds used in the MCMC process for each model parameter.}
	 	\label{priors}
	 \end{table}

\begin{figure}
	 	\centering
	 	\includegraphics[scale = 0.4]{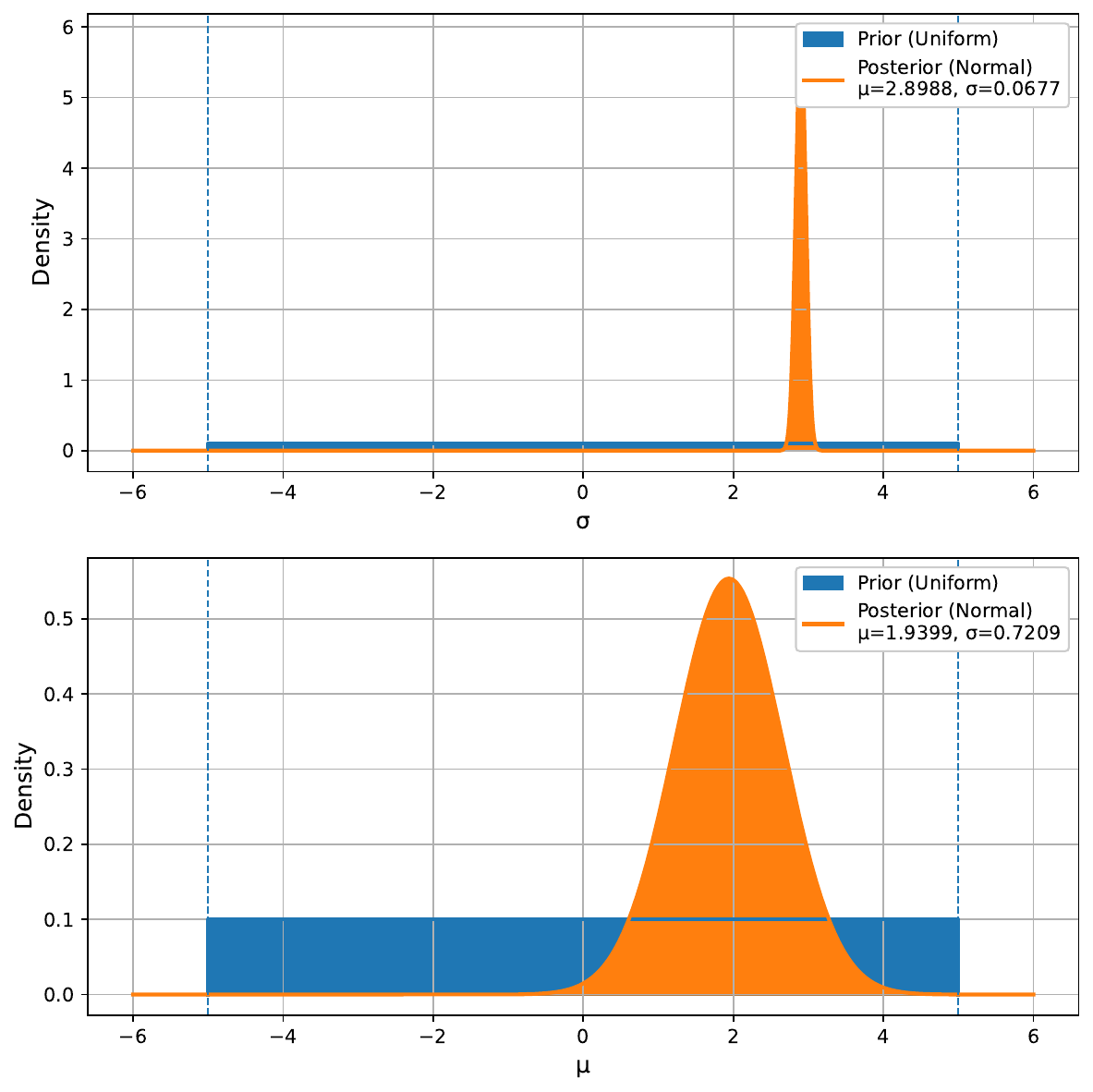}
	 	\caption{ Posterior distributions of the $\sigma$ and $\mu$ parameters overlaid on the prior range for model \texttt{M3VEXP}.}
	 	\label{priorposterior}
	 \end{figure}

	 To make sure that our prior choice does not influence the results, we analysed the prior-posterior relationship for each model and show a representative plot in Fig.~\ref{priorposterior}. Specifically, we compare the posterior distribution that resulted from the MCMC run and the intial prior range for model \texttt{M2VEXP}, proving that the walkers clearly converge independently of the prior distribution.
	 
	\subsection{Goodness-of-fit Statistical Analysis}
	
	In the following, we analyse the goodness-of-fit that resulted from the MCMC process by proposing a reduced $\chi^2$ statistic given by,
	\begin{equation}
		\chi^2_{red} = \frac{\chi^2}{\nu},
	\end{equation}
	where $\nu$ represent the degrees of freedom of the models, represented by the number of data points, which in our case is 3, coming from the 3 considered nuclear abundances, minus the number of fitted parameters.  For the case of the quadratic potential we have $\nu = 3 - 1 = 2$ degrees of freedom and for the Higgs type and exponential potential, $\nu = 3 - 2 = 1$, respectively. 
	
	The probability of exceeding the $\chi^2$ value in a measurement is given by \cite{stat},
	\begin{equation}
		P_\chi(\chi^2; \nu) = \int_{\chi^2}^{\infty} \frac{1}{2^{\nu/2} \Gamma(\nu/2)} (x^2)^{(\nu-2)/2} e^{-x^2/2} \, d(x^2), \label{probability}
	\end{equation}
	where $P_\chi(\chi^2; \nu) = 0.5$ indicates that the model residuals are consistent with Gaussian-distributed noise around the mean, with no significant over- or under-fitting. 
	For our models, this probability is achieved for $\chi^2_{red} = 0.445$ for $\nu = 1$ and $\chi^2_{red} = 0.693$ for $\nu = 2$.  For higher $\nu$ the $\chi^2_{red}$ statistic approaches the expected value of 1.
	
	 At most times, $\chi^2_{red}$ is a convenient indicator for describing the discrepancies between the variance of the fit and the variance of the data. However, its sampling distribution depends explicitly on $\nu$ and it can appear broad and skewed when the number of degrees of freedom is small. Therefore, the $p$-value becomes a more reliable metric as the $P_\chi(\chi^2;\nu)$ statistic uses a tail probability of the $\chi^2$ distribution that is directly comparable across models with low-DOF. For this reason, we treat $P_\chi$ as the primary goodness-of-fit metric and report $\chi^2_{red}$ only as an intermediary result, as relying on the $\chi^2_{red} \approx 1$ statistic alone can be misleading in small-sample sizes with non-linear settings.

	To further evaluate the model support based on complexity and fit, we construct statistical metrics such as Akaike Information Criterion and Bayesian Information Criterion \cite{aic1, bic1} based on the maximum value of the log likelihood function. The $\mathrm{AIC}$ statistics is given by,
	\begin{equation}
		\mathrm{AIC} = -2\log L_{max} + 2k,
	\end{equation}
	where $k$ denotes the total number of free parameters in the model. Likewise, the $\mathrm{BIC}$ statistics is given by the following expression,
	\begin{equation}
		\mathrm{BIC} = -2 \log L_{max} + k \log n,
	\end{equation}
	where $n$ is the number of data points. We introduce the relative differences
	\begin{equation}
		\Delta_{\mathrm{AIC}} = \mathrm{AIC}_i - \mathrm{AIC}_{\min},
	\end{equation}
	\begin{equation}
		\Delta_{\mathrm{BIC}} = \mathrm{BIC}_i - \mathrm{BIC}_{\min},
	\end{equation}
	where $\mathrm{AIC}_{\min}$ and $\mathrm{BIC}_{\min}$ represent the lowest $\mathrm{AIC}$ and $\mathrm{BIC}$ values among all tested models. The closer these differences are to zero, the better the goodness of fit, such that if $\Delta_{\mathrm{AIC}} \leq 2$ and $\Delta_{\mathrm{BIC}} \leq 2$, the model has strong evidence in its favour, if $4 \leq \Delta_{\mathrm{AIC}} \leq 7$ and $2 \leq \Delta_{\mathrm{BIC}} \leq 6$, the model is less supported and if $\Delta_{\mathrm{AIC}} > 10$ and $\Delta_{\mathrm{BIC}} > 6$, there is little to no support for the model \cite{aic2}.
	
	Table \ref{results} presents the $\chi^2_{\mathrm{red}}$ values for all nine cosmological models, providing a first measure of the goodness-of-fit. For quadratic potential models having $\nu = 2$, model \texttt{M1VPHI2} has $\chi^2_{red} = 0.6816$, corresponding to $P_\chi(\chi^2; 2) \approx 0.5$, which indicates an ideal statistical fit. The second and third models  \texttt{M2VPHI2} and \texttt{M3VPHI2}  yield slightly lower $\chi^2_{red}$ values, resulting in $P_\chi(\chi^2; 2) = 0.7$. This suggests that the data fluctuations are smaller than expected, which could indicate slight model overfitting, but still within acceptable limits.

			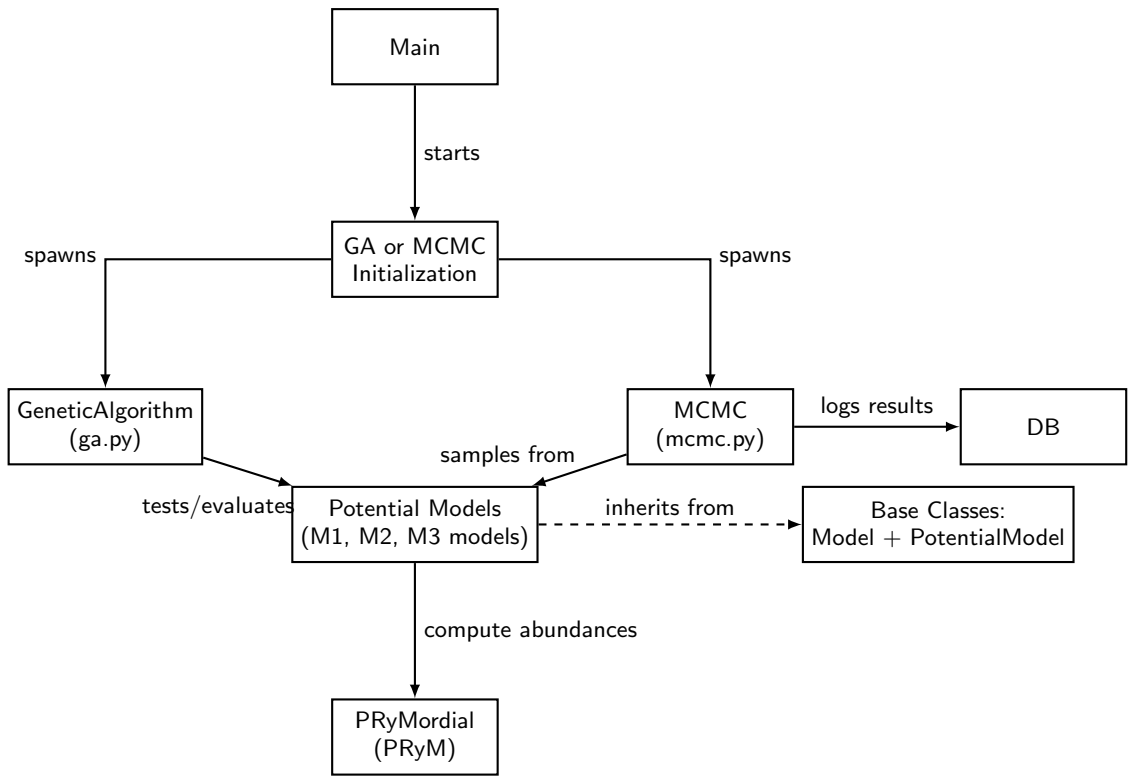
\begin{figure*}[htbp!]
		\centering
		\begin{tikzpicture}[
			font=\sffamily,
			box/.style={draw, thick, minimum width=2.2cm, minimum height=1cm, align=center},
			arrow/.style={->, thick, >=latex},
			dashedarrow/.style={->, thick, dashed, >=latex},
			node distance=1.8cm and 2.2cm
			]
			
			\node[box] (main) {Main};
			
			\node[box, below=of main] (threads) {GA or MCMC \\ Initialization};
			
			\node[box, below left=1.2cm and 1.7cm of threads] (GA) {GeneticAlgorithm\\(ga.py)};
			\node[box, below right=1.2cm and 1.7cm of threads] (MCMC) {MCMC\\(mcmc.py)};
			
			\node[box, right=of MCMC] (DB) {DB};
			
			\node[box, below=2.5cm of threads] (models) {Potential Models\\(M1, M2, M3 models)};
			
			\node[box, below=of models] (prym) {PRyMordial\\(PRyM)};
			
			\node[box, right=3.5cm of models] (base) {Base Classes:\\Model + PotentialModel};
			
			
			\draw[arrow] (main) -- node[midway,right]{starts} (threads);
			
			\draw[arrow] (threads) -| node[pos=0.5,left]{spawns} (GA);
			\draw[arrow] (threads) -| node[pos=0.5,right]{spawns} (MCMC);
			
			\draw[arrow] (MCMC) -- node[midway,above]{logs results} (DB);
			
			\draw[arrow] (GA) -- node[midway,above left=-30pt]{tests/evaluates} (models);
			
			\draw[arrow] (MCMC) -- node[midway,above left=-3pt]{samples from} (models);
			
			\draw[arrow] (models) -- node[midway,right]{compute  abundances} (prym);
			
			\draw[dashedarrow] (models.east) to[out=0,in=180] 
			node[midway,above]{inherits from} (base.west);
			
		\end{tikzpicture}
		\caption{Schematic representation of the \texttt{genesys} program.}
		\label{genesys}
	\end{figure*}

	In the case of Higgs-type and exponential potentials having one degree of freedom, most models have  $P_\chi(\chi^2; 1) \approx 0.4$. This implies that \texttt{M2VPHI24}, \texttt{M2VEXP}, \texttt{M3VPHI24} and \texttt{M3VEXP} models are slightly less precise in describing the data. In contrast, models \texttt{M1VPHI24} and \texttt{M1VEXP} have lower probability $P_\chi(\chi^2; 1) \approx 0.25$, suggesting their chi-square values are relatively high so that the models may fail to capture the data behaviour.

	To complement this analysis, we use the Akaike and Bayesian information criteria, which evaluate model quality while penalizing complexity. For each potential type, we identify the model with the minimum AIC and BIC values and compare others against it.

	For quadratic potentials, the best model is \texttt{M2VPHI2}, with $\mathrm{AIC}_{\min} = 2.6768$ and $\mathrm{BIC}_{\min} = 1.7726$. Model \texttt{M1VPHI2} has  $\Delta_{\mathrm{AIC}} = 0.6864$ and $\Delta_{\mathrm{BIC}} = 0.6892$, suggesting substantial support. Model \texttt{M3VPHI2} shows even smaller differences, $\Delta_{\mathrm{AIC}} = 0.0661$ and $\Delta_{\mathrm{BIC}} = 0.0689$, indicating a nearly equal performance.
	
	Among Higgs-type potentials, model \texttt{M2VPHI24} achieves the minimum $\mathrm{AIC}_{\min} = 4.6802$  and $\mathrm{BIC}_{\min} = 2.8777$. Both \texttt{M1VPHI24} and \texttt{M3VPHI24} have low $\Delta_{\mathrm{AIC}}$ and $\Delta_{\mathrm{BIC}}$ less than 1, indicating good support.
	
	Similarly, for exponential potentials, model \texttt{M2VEXP} minimizes both criteria, $\mathrm{AIC}_{\min} = 2.6920$ and $\mathrm{BIC}_{\min} = 2.8393$. The third model \texttt{M3VEXP} has $\Delta_{\mathrm{AIC}} = 0.2054$ and $\Delta_{\mathrm{BIC}} = 0.055$, showing string support, while \texttt{M1VEXP} shows moderate support with higher $\Delta_{\mathrm{BIC}} = 0.7697$.
	
	\subsection{The \texttt{genesys} code for BBN constraints on cosmological models}
	
	The methodology presented in this section was synthesized into a python code named \texttt{genesys}, whose implementation structure can be observed in Fig~\ref{genesys}. This program offers two primary workflows: a Genetic Algorithm mode and an MCMC mode, depending on the user's objective.
	
	The Genetic Algorithm mode is designed to determine initial parameter values for the cosmological models, starting with the null potential case, the simplest model configuration. In this setup, the scalar potential is assumed to vanish, allowing us to isolate the behaviour of the dynamical system governed only by the background equations. The goal here is to find parameter values that are consistent with solutions to the differential system in this minimal scenario. Once these initial conditions are obtained, they can later be used in more complex models involving quadratic, Higgs, or exponential scalar potentials.
	
	To initiate a GA run, the user selects a specific null potential model and provides algorithm-specific settings such as the numerical solver, parameter interval and tolerance thresholds. The code then follows the evolutionary steps illustrated in Fig.~\ref{genesys}, ultimately reporting the best-fit parameter set to the console.
	
	Alternatively, the user can choose to perform an MCMC analysis, used to statistically constrain the parameters of non-null scalar potential models by comparing their predictions to observed primordial abundances. Here, the user must specify the potential model, the number of MCMC steps, the number of walkers, and the prior bounds for each parameter. 
	
	During the MCMC process, each walker proposes a set of parameters within a specified model, then attempts to compute the associated additional energy density and pressure contributions from the scalar fields. If these computations fail due to numerical instability or unphysical results, the walker is immediately rejected. Otherwise, the model proceeds to evaluate the BBN abundances and compute the corresponding $\chi^2$.  Finally, if all calculations were successfully executed, the parameters and abundances are saved in the database for further statistical analysis.

	At the core of the \texttt{genesys} library is the abstract \texttt{Model} class, which exposes an  interface for both GA and MCMC compatible models. Derived from it, the abstract \texttt{PotentialModel} class specializes this interface for cosmological scalar field models, providing internal methods for solving the system of ordinary differential equations and specifying the scalar potential.
	
	Concrete implementations begin with the null potential models, which inherit from \texttt{PotentialModel} class. These base classes are then extended to involve non-null scalar potential models, which are constrained using observational data through the PRyMordial framework.

	The \texttt{PotentialModel} class reintegrates the modified Friedmann equations, which account for radiation non-conservation that arises due to the matter coupling with the additional scalar fields. Each model instance solves the coupled system of ODEs given in Eq.~\ref{t1}-\ref{t12}
	over a fixed time interval using the \texttt{scipy.integrate.solve\_ivp} method with adaptive step sizing.
	
	The time-temperature relation is computed through the \texttt{\_compute\_time\_map()} method, which integrates the Hubble solution to obtain
	\begin{equation}
		T(t) = T_0 \exp\left(-\int_0^t H(\tau)\,d\tau\right),
	\end{equation}
	where $T_0 = 10$ MeV is the temperature assigned at the beginning of the simulation. This integration is performed using \texttt{scipy.integrate.cumulative\_trapezoid} over the time domain $t \in [0, 10^7]$ s to ensure that the entire BBN epoch is covered.
	
	\begin{figure*}[htbp!]
		\centering
		\includegraphics[scale = 0.66]{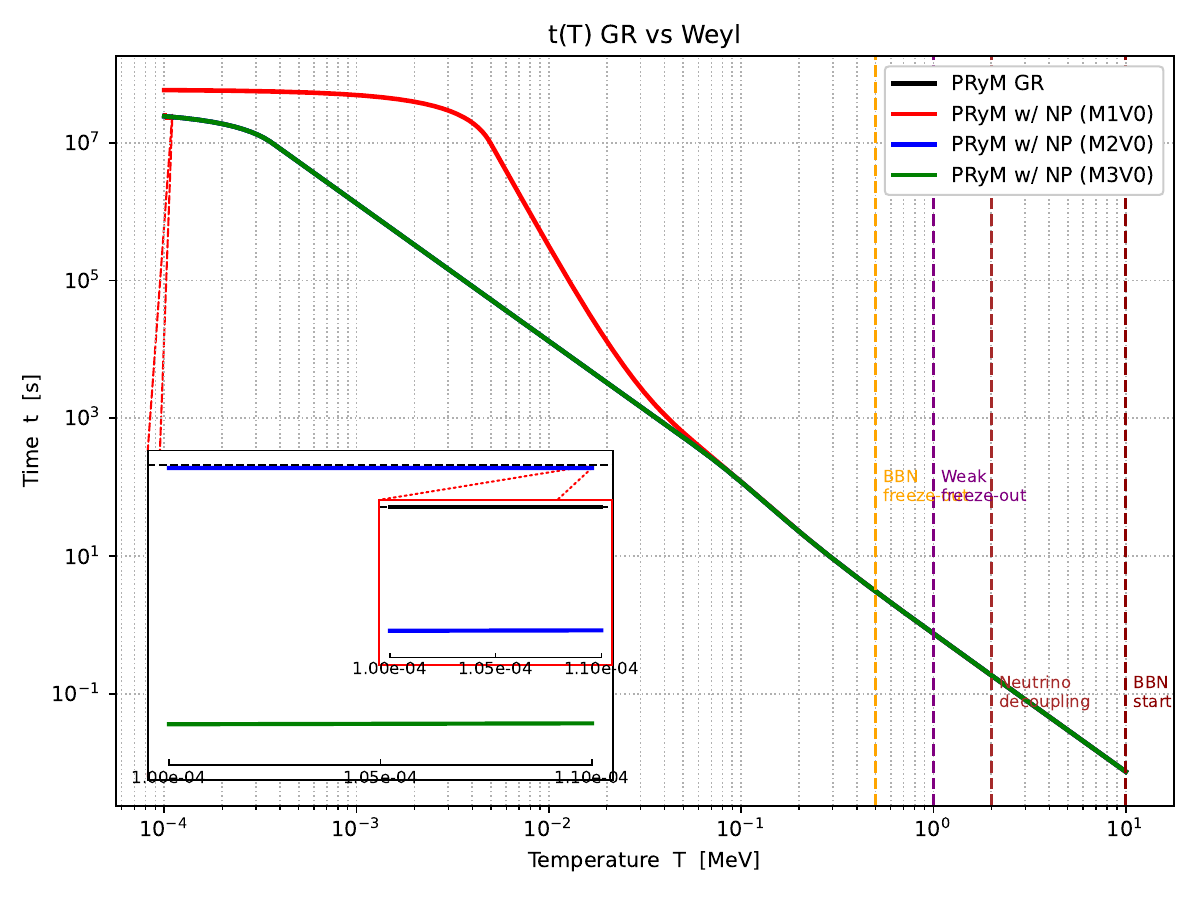}
		\caption{ Time evolution as a function of temperature represented on a logarithmic scale for the epoch considered in PRyMordial. The main plot shows the overall $t(T)$ relationship, comparing GR and Weylian models (M1V0, M2V0, M3V0). Vertical dashed lines mark key BBN events: freeze-out (0.5 MeV), weak interaction freeze-out (1.0 MeV), neutrino decoupling (2.0 MeV), and BBN start (10.0 MeV).}
		\label{tofT}
	\end{figure*}
	
	The modified thermodynamic functions are directly computed in \texttt{\_compute\_functions\_time()}, which maps the time evolution to the temperature evolution by using the \texttt{tofT} function in \texttt{PRyMordial}, such that the total energy density and pressure are computed,
	\begin{align}
		\rho_{\text{tot}}(T) &= \rho_{\text{rad}}(T) + \rho_w(T)\\
		p_{\text{tot}}(T) &= p_{\text{rad}}(T) + p_w(T).
	\end{align}
	
	These geometric functions $\rho_w$ and $p_w$ are dimensionalized and passed to the \texttt{PRyMclass} constructor through the New Physics (NP) parameters \texttt{rho\_NP}, \texttt{p\_NP}, and \texttt{drho\_NP\_dT}. This ensures that all nuclear reaction rates, thermal equilibrium conditions, and freeze-out period operate within the same framework. The bidirectional mapping \texttt{tofT} and \texttt{Toft} allows for direct integration with PRyMordial's nucleosynthesis calculations, by making use of the modified cosmological evolution given by the Hubble parameter.
	
	To observe the differences between the GR and Weylian time evolution as a function of temperature, we plotted in Fig.~\ref{tofT} three instances of the Weylian models along with the standard GR evolution, in which $T \propto t^{-1/2}$. The resulting time-temperature relations $t_{\text{GR}}(T)$ and $t_{\text{NP}}(T)$ display a  similar behaviour, with relative deviations $\Delta t/t_{\text{GR}} < 0.1\%$ across the BBN temperature range, from 0.1 keV to 10 MeV. A small shift can be observed for \texttt{M1V0} in which both scalar fields couple to matter, especially for a temperature at which the nucleosynthesis is in progress. The other two models \texttt{M2V0} and \texttt{M3V0} have very small deviations due to the the coupling parameter $\alpha^2$, which becomes much smaller compared to the contribution for the first model. Interestingly enough, the second model's $t(T)$ function appears to be virtually identical to the GR limit.

	\section{Discussions and Final Remarks} \label{concl}

	The present work investigated three cosmological  models, which include a boundary influence present in the early Universe,  by imposing relevant constraints provided by the Big Bang Nucleosynthesis theory. The light nuclei abundances in the form of hydrogen, deuterium, helium-3, helium-4 and lithium-7 can assert the viability of the cosmological frameworks, based on the model's ability to predict realistic behaviour for the nucleosynthesis process.
	
	The present study extends the models introduced in \cite{warmweyl} beyond the inflationary period, and explores the contribution of the boundary terms during the BBN era. While the intensive production of radiation by two scalar fields ${-}$ the inflaton and the Weyl boundary terms ${-}$ is not the dominant physical process during the BBN phase, a limited creation of matter and radiation cannot be excluded even at this relatively late stage in the Universe's evolution.

In the present gravitational theory which also includes the effects of a boundary,  the dissipative scalar field $\phi$ as well as the second scalar field $\psi$, coming from the Weyl vector, are still responsible for the dynamical evolution of the Universe after the particle creation era,  described as a warm inflationary period. The two scalar fields interact dynamically with the matter, as can be seen from the energy balance equation Eq.~(\ref{cons_eq}). On the other hand the dynamics of the scalar field, as described by the Klein-Gordon equation  Eq.~(\ref{KG}), is determined only by the self-interaction potential of the field, and by the dissipative effects. 

The general conservation equation we have used in our investigations was initially obtained in \cite{warmweyl}, by considering the modified gravitational field equations Eq.~(\ref{fe}), which resulted by considering the influence of the Weyl boundary, characterized by the vector $\omega_\mu$. In the approach adopted in the present work we have assumed that the Weyl vector is the gradient of a scalar function $\psi$, and thus we have considered the boundary of the space-time as described by an integrable Weyl geometry. In this approach extends the boundary is Weylian, and the transition to a Weylian boundary is realized by substituting the Riemannian covariant derivatives with their  Weyl geometric counterparts. 

The gravitational field equations that include boundary terms were considered for  a Friedman-Lema\^{i}tre-Robertson-Walker metric, which led to the  modified Friedmann equations Eq.~(\ref{F1}, \ref{F2}). These equations, together with the dissipative Klein-Gordon equation and the global conservation equation, were used to construct three different cosmological scenarios that describe the early evolution of the Universe. The first model decoupled the evolution of the Weyl scalar field from the other cosmological constituents, while adopting a nonconservative form for the matter sector, with the extra terms, depending on the cosmological scalar field and on the boundary term could also contribute to matter creation even in this later stage of cosmological evolution. In the second model the nonconservative matter evolution is determined  only by the scalar field $\phi$, whereas in the third proposed scenario the matter evolution is influenced entirely by the dynamics of the Weyl vector. 
	
The methods used for constraining the previously presented theoretical models rely on the work of \cite{Capozziello, fQ}, where the $f(T)$ teleparallel gravity, as well as $f(Q)$ gravity were constrained by the measurement of helium-4 abundance, as indicated by the primordial mass fraction estimate $Y_P$, due to the variation caused by the associated deviation in the freezing temperature $\delta T_f/T_f$. Within our theoretical framework, the constraint provided by the freeze-out temperature deviations is given by Eq.~(\ref{upper}).

\begin{figure*}
 		\centering
 		\includegraphics[scale = 0.55]{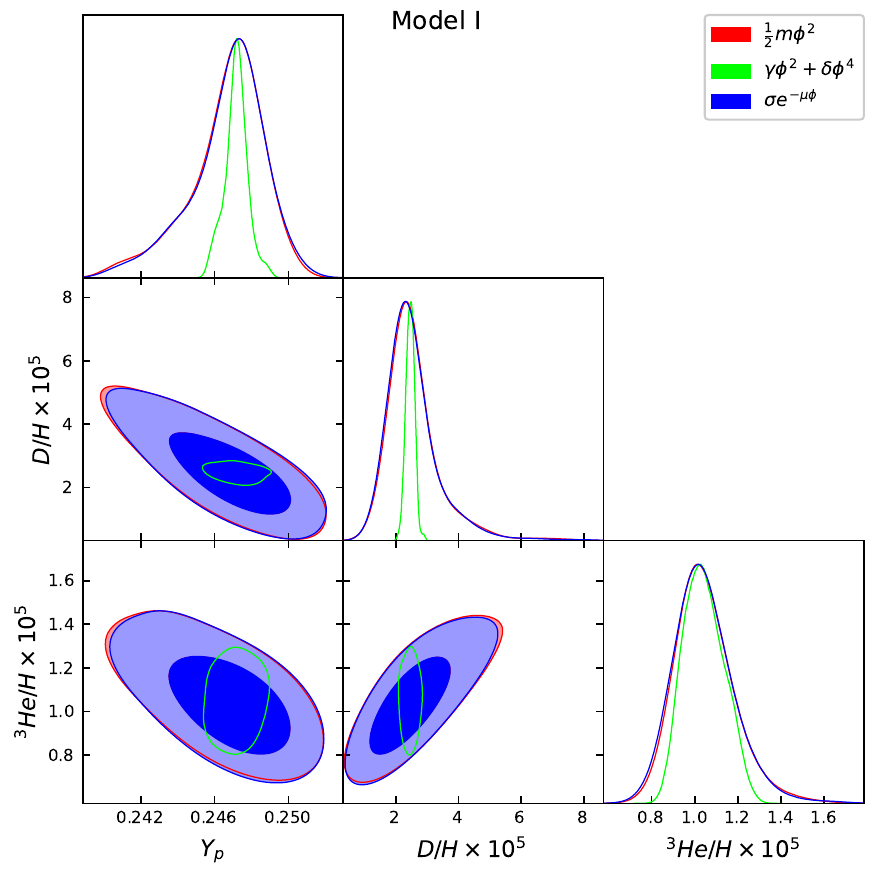}
 		\includegraphics[scale = 0.55]{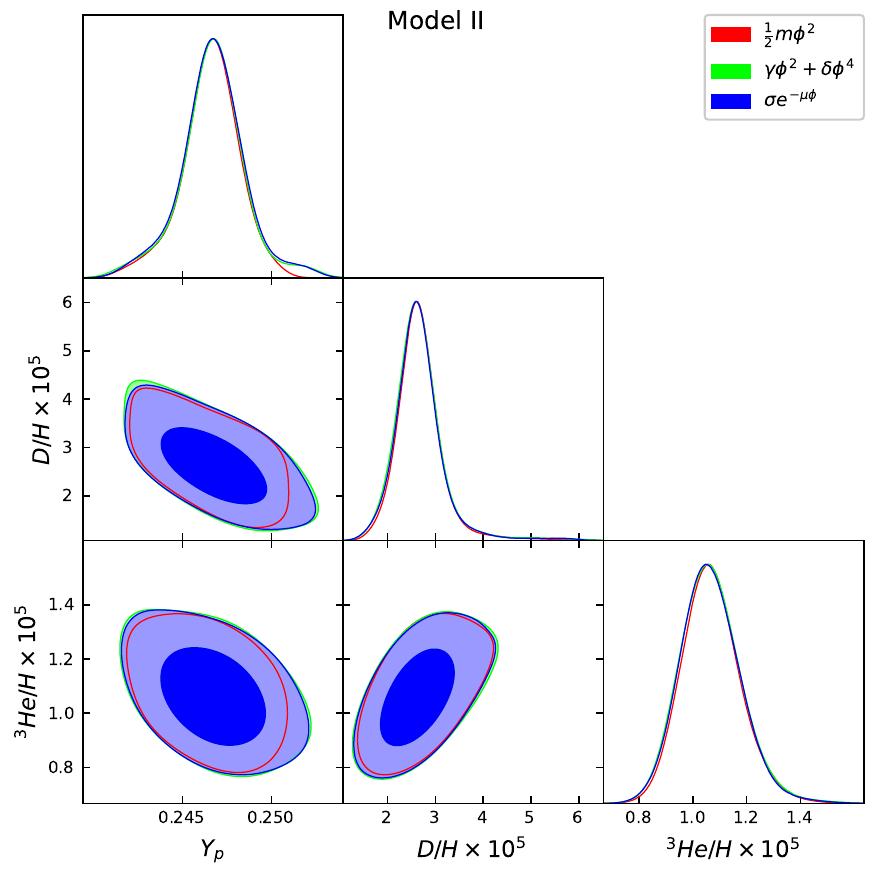}
 		\includegraphics[scale = 0.55]{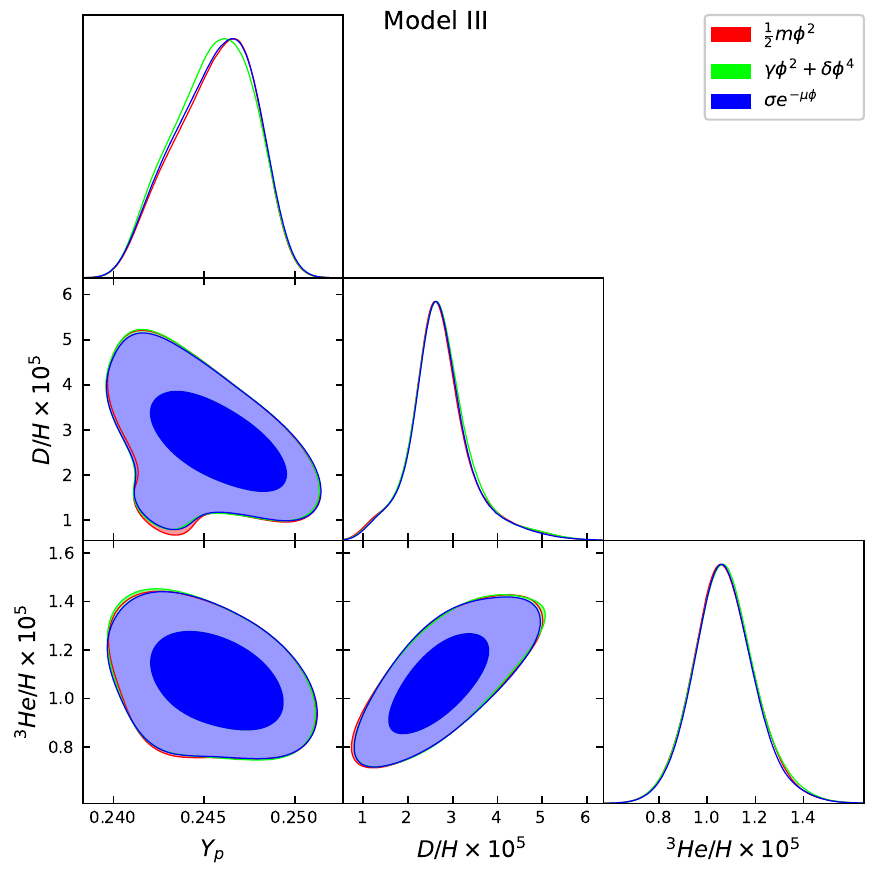}
 		\caption{Light element abundances for the each model (first - upper left, second - upper right, third - bottom) with the three scalar potential types overlaid.}
 		\label{corners}
 	\end{figure*}

 Our work has been implemented in the python program \texttt{genesys}, whose schematic is given in Fig.~\ref{genesys}, which supports both Genetic Algorithm and MCMC parameter estimation. It interfaces \texttt{Potential Models} objects with the \texttt{PRyMordial} library to compute BBN abundances, enabling efficient exploration of scalar potential parameters. Using the results  provided in Fig.~\ref{results_potentials}, we implemented a Monte Carlo analysis to generate the corner plots illustrated in Figure~\ref{corners}. The approach involves fixing the scalar potential parameters for each model, while sampling the nuclear reaction rates, the neutron mean lifetime $\tau_n$ and the baryon to photon ratio $\Omega_{b} h^2$ from their respective probability distributions,
 \begin{eqnarray}
 	\tau_n &=& 879.4 \pm 0.5 \; s, \\
 	\Omega_{b} h^2 &=& 0.0223 \pm 0.0002,
 \end{eqnarray}
 within a 1$\sigma$ confidence level.

 \begin{table*}[htbp!]
 	\centering
 	
 	\renewcommand{\arraystretch}{1.3}
 	\setlength{\tabcolsep}{10pt}
 	\begin{tabular}{|l|l|c|}
 		\hline
 		Model & Description & Parameters ($\mu \pm \sigma$) \\
 		\hline \hline
 		\multirow{9}{*}{Model I} 
 		& $v(\phi) = \tfrac{1}{2}m\phi^2$ 
 		& $m = 0.00112 \pm 0.00032$ \\
 		& $v(\phi) = \gamma \phi^2 + \delta \phi^4$ 
 		& $\gamma = -0.00551 \pm 0.00239$, \quad $\delta = 2.19851 \pm 0.95843$ \\
 		& $v(\phi) = \sigma e^{-\mu \phi}$ 
 		& $\sigma = 2.82453 \pm 0.05340$, \quad $\mu = 2.51592 \pm 1.18241$ \\
 		\cline{2-3}
 		& Initial conditions 
 		& $\phi_0 = -0.0261 \pm 0.0271$, \quad $\dot{\phi}_0 = -0.3796 \pm 0.4159$ \\
 		&  
 		& $\psi_0 = -0.2597 \pm 1.9600$, \quad $\dot{\psi}_0 = 0.0738 \pm 1.9600$ \\
 		\cline{2-3}
 		& Weyl coupling 
 		& $\alpha = -0.1899 \pm 0.1009$ \\
 		& Conservation parameter 
 		& $\lambda = 0.5648 \pm 0.4532$ \\
 		& Dissipation coefficient 
 		& $\beta = -142.82 $ \\
 		\hline \hline
 		\multirow{9}{*}{Model II} 
 		& $v(\phi) = \tfrac{1}{2}m\phi^2$ 
 		& $m = 0.00048 \pm 0.00018$ \\
 		& $v(\phi) = \gamma \phi^2 + \delta \phi^4$ 
 		& $\gamma = -0.00041 \pm 0.00020$, \quad $\delta = 0.32267 \pm 0.12905$ \\
 		& $v(\phi) = \sigma e^{-\mu \phi}$ 
 		& $\sigma = 2.89497 \pm 0.05128$, \quad $\mu = 1.56283 \pm 0.69417$ \\
 		\cline{2-3}
 		& Initial conditions 
 		& $\phi_0 = -0.0369 \pm 1.8155$, \quad $\dot{\phi}_0 = -0.0174 \pm 1.0071$ \\
 		&  
 		& $\psi_0 = 0.0495 \pm 1.7080$, \quad $\dot{\psi}_0 = -0.0475 \pm 0.0803$ \\
 		&  
 		& $\psi_{02} = 0.0373 \pm 0.1878$ \\
 		\cline{2-3}
 		& Weyl coupling 
 		& $\alpha = 0.0239 \pm 0.0784$ \\
 		& Dissipation coefficient 
 		& $\beta = -268.13$ \\
 		\hline \hline
 		\multirow{9}{*}{Model III} 
 		& $v(\phi) = \tfrac{1}{2}m\phi^2$ 
 		& $m = 0.00062 \pm 0.00022$ \\
 		& $v(\phi) = \gamma \phi^2 + \delta \phi^4$ 
 		& $\gamma = -0.00003 \pm 0.00002$, \quad $\delta = 0.58382 \pm 0.20160$ \\
 		& $v(\phi) = \sigma e^{-\mu \phi}$ 
 		& $\sigma = 2.97062 \pm 0.01394$, \quad $\mu = 1.69332 \pm 0.63128$ \\
 		\cline{2-3}
 		& Initial conditions 
 		& $\phi_0 = -0.0094 \pm 0.1933$, \quad $\dot{\phi}_0 = -0.0009 \pm 0.0003$ \\
 		&  
 		& $\psi_0 = -0.0880 \pm 0.1500$, \quad $\dot{\psi}_0 = 0.0155 \pm 0.0330$ \\
 		&  
 		& $\psi_{02} = -0.0495 \pm 0.1030$ \\
 		\cline{2-3}
 		& Weyl coupling 
 		& $\alpha = 0.0233 \pm 0.0247$ \\
 		& Dissipation coefficient 
 		& $\beta = -1682.27 $ \\
 		\hline
 	\end{tabular}
 	\caption{ Summary of mean values and standard deviations ($\mu \pm \sigma$) for all model parameters, including derived $\beta$ values calculated from the analytical expression.}
 	\label{potentials}
 \end{table*}

 We present the results for the model parameters form both the MCMC analysis and the Genetic Algorithm approach from Section \ref{alphapy} in Table~\ref{potentials}.

	 \begin{figure*}
	 	\centering
	 	\includegraphics[scale = 0.35]{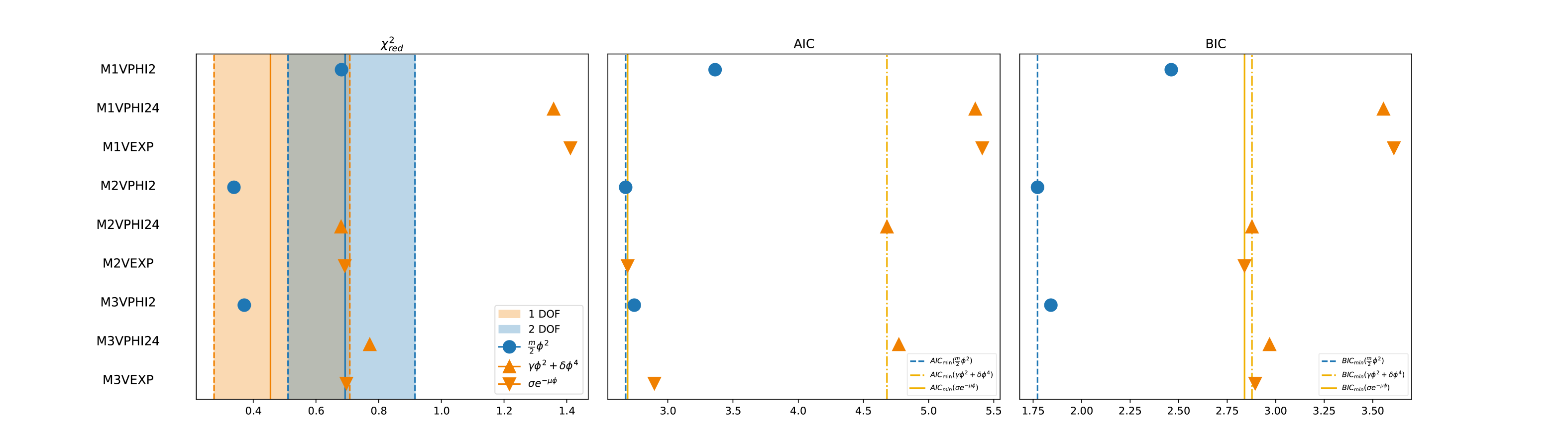}
	 	\caption{Visual representation of the model performance based on $\chi^2_{red}$, AIC and BIC statistics. The coloured areas (left panel) represent $\chi^2_{red}$ values within 0.6-0.4 probabilities $P_{\chi}(\chi^2, \nu)$ (left dashed line $P_\chi = 0.6$, middle solid line $P_\chi = 0.5$, right dashed line $P_\chi = 0.4$). The AIC$_{\rm min}$ values (middle panel) are marked by lines for each potential type and likewise for the BIC$_{\rm min}$ values (right panel). The symbols are shown alongside each potential model, listed on the left.}
	 	\label{ranking}
	 \end{figure*}

	 The results of our  statistical  analysis are summarized in Fig~\ref{ranking} based on which we note that all three quadratic potentials models with $\nu = 2$ (blue shaded area for the \texttt{PHI2} potential classes) present $\chi^2_{red}$ values around $0.34$-$0.68$ with probabilities of $0.5$ to $0.7$, indicating a generally good fit to the observational data. Among these, \texttt{M1VPHI2} provides an ideal fit quality, while \texttt{M2VPHI2} and \texttt{M3VPHI2} show slightly higher probabilities of exceeding the chi-squared value (lower $\chi^2_{red}$), suggesting marginal overfitting within acceptable bounds. The AIC and BIC metrics further confirm that \texttt{M2VPHI2} achieves the best overall performance, followed by \texttt{M3VPHI2} and \texttt{M1VPHI2}, all of which have statistical support.

	 For the Higgs-type and exponential potentials with $\nu=1$ (orange shaded area for the \texttt{PHI24, EXP} potential classes), the fits are somewhat less precise, with probabilities between $0.25$ and $0.4$, suggesting underfitting. Despite this, AIC and BIC comparisons reveal \texttt{M2VPHI24} and \texttt{M2VEXP} as the most favourable models in their respective potential classes. Both show low $\Delta_{\mathrm{AIC}}$ and $\Delta_{\mathrm{BIC}}$ values, indicating a strong support relative to other models. In contrast, higher $\Delta_{\mathrm{AIC}}$ and $\Delta_{\mathrm{BIC}}$ values were obtained for \texttt{M1VPHI24} and \texttt{M1VEXP}, suggesting less support from the data.
	 
	 In terms of primordial abundances, all models predict $Y_p \simeq 0.246$ and $^3\rm{He/H} \simeq 1.046\times10^{-5}$ within narrow uncertainties, with slight variations in D/H. From the corner plots in Fig.~\ref{corners}, we observe a strong negative correlation between the $Y_p$ and D/H and between $Y_p$ and $^3$He/H, respectively. There also exists a positive correlation between the D/H and $^3$He/H abundances. More importantly, all models are in agreement with each other, aside from \texttt{M1VPHI24} which  shows a relatively poorer fit, a conclusion which can be also observed in the $\chi^2_{red}$ value. Finally, it can be concluded that the three models predict the primordial nuclear abundances given by the SBBN. 
	 
	 In addition to the visual representation provided in Fig.~\ref{ranking}, we rank model performance based on the $\chi^2_{red}$ value with its associated probability of exceeding it Eq.~\ref{probability} and on the AIC and BIC coefficients (Table \ref{performance}). Since the potential classes \texttt{(VPHI2, VPHI24, VEXP)} are evaluated within their respective categories of scalar field potentials, we compare performance across the three cosmological models. This comparison is necessary because the quadratic potential class involves a different number of degrees of freedom ($\nu = 2$) compared to the Higgs-type and exponential potentials ($\nu = 1$). Consequently, we distinguish between the Higgs-type and exponential classes in our statistical ranking.
	 In Table~\ref{performance}, we identify \texttt{M2VPHI24} as the best-performing model within the Higgs-type potential class, \texttt{M2VEXP} for the exponential class, and \texttt{M2VPHI2} for the quadratic potentials, as each have the lowest AIC and BIC coefficients within its respective potential class.

	 We emphasize that no single statistical measure fully captures the fit quality while accounting for model complexity. In particular, for the low number of degrees of freedom involved, we find the p-value associated with $\chi^2_{red}$ to be a reliable first indicator of goodness-of-fit, especially when $\nu=1$ or $2$. However, model selection based only on p-values can be misleading. Table~\ref{performance} and the posteriors in Fig.~\ref{corners} show that even models with better p-values can be less supported due to higher penalties in AIC/BIC or when their nucleosynthesis predictions are narrower than expected.
	 
	 A concrete example is the model \texttt{M1VEXP}, which shows better nuclear abundance predictions and lower $\Delta$AIC/BIC than \texttt{M1VPHI24}, which has a slightly higher p-value, thus being the lowest ranked in Table~\ref{performance}. Therefore we classify model performance based on a multi-metric evaluation, as the overall interpretation of a model should also rely on other physical predictions, such as light element abundances, besides the associated statistical results.

	 \begin{table}
	 	\centering
	 	\renewcommand{\arraystretch}{1.4}
	 	\begin{tabular}{|c|l|}
	 		\hline
	 		Model & Observations \\
	 		\hline
	 		\texttt{M1VPHI2} & $P_\chi(\chi^2; 2) = 0.506$, $\Delta_{\mathrm{AIC}} = 0.686$, $\Delta_{\mathrm{BIC}} = 0.689$ \\
	 		\hline
	 		\texttt{M2VPHI24} & $P_\chi(\chi^2; 1) = 0.409$, AIC$_{\rm min}$ \& BIC$_{\rm min}$ (\texttt{VPHI24})\\
	 		\hline 
	 		\texttt{M2VEXP} & $P_\chi(\chi^2; 1) = 0.405$, AIC$_{\rm min}$ \& BIC$_{\rm min}$ (\texttt{VEXP})\\
	 		\hline 
	 		\texttt{M3VEXP} & $P_\chi(\chi^2; 1) = 0.403$, $\Delta_{\mathrm{AIC}} = 0.205$, $\Delta_{\mathrm{BIC}} = 0.05$ \\
	 		\hline 
	 		\texttt{M3VPHI24} & $P_\chi(\chi^2; 1) = 0.379$, $\Delta_{\mathrm{AIC}} = 0.092$, $\Delta_{\mathrm{BIC}} = 0.091$\\
	 		\hline 
	 		\texttt{M2VPHI2} & $P_\chi(\chi^2; 2) = 0.712$, AIC$_{\rm min}$ \& BIC$_{\rm min}$ (\texttt{VPHI2})\\
	 		\hline 
	 		\texttt{M3VPHI2} & $P_\chi(\chi^2; 2) = 0.689$, $\Delta_{\mathrm{AIC}} = 0.066$, $\Delta_{\mathrm{BIC}} = 0.069$\\
	 		\hline 
	 		\texttt{M1VEXP} & $P_\chi(\chi^2; 1) = 0.234$, $\Delta_{\mathrm{AIC}} = 0.205$, $\Delta_{\mathrm{BIC}} = 0.055$\\
	 		\hline 
	 		 \texttt{M1VPHI24} &  $P_\chi(\chi^2; 1) = 0.244$, $\Delta_{\mathrm{AIC}} = 0.678$, $\Delta_{\mathrm{BIC}} = 0.678$\\
	 		\hline 
	 	\end{tabular}
	 	\caption{Model ranking from the best (first) to the worst (last) based on the $\chi^2$, AIC and BIC statistics.}
	 	\label{performance}
	 \end{table}
	 
	 The results support models where the cosmological dynamics is mainly determined  by the evolution of the scalar field $\phi$, which is characteristic to the second considered cosmological scenario. Specifically, the quadratic model \texttt{M2VPHI2} and the exponential model \texttt{M2VEXP} show the best AIC and BIC values, along with good $\chi^2$ agreement with BBN abundances. Notably, \texttt{M1VPHI2}, which includes both inflaton dynamics and Weyl boundary effects, provides the best overall fit to the data, with the lowest $\chi^2_{red}$, suggesting that boundary-influenced cosmological evolution could provide a viable mechanism for explaining the physical characteristics of the early Universe.

 We would like to point out that in the present approach we have adopted an approach based on the Weyl Integrable geometry, by assuming that the vector field $\omega _\mu$ is the gradient of a scalar field, so that $\omega _\mu =\nabla _\mu \psi$. Therefore, the boundary contribution of the Weyl vector adds a Brans-Dicke type scalar field contribution to the field equations, represented by a kinetic term only, but without an associated potential. Moreover,  the theoretical model also includes a standard scalar field $\phi$, with a self-interaction potential $V(\phi)$. The two scalar fields do not interact, and there is no non-minimal coupling between the scalar fields and the gravitational sector. One of the important problems facing gravitational theories is the presence of various types of instabilities \cite{In1,In2,In3,In4,In5}.  The ghost instabilities do appear generally due to the wrong (negative) sign of the kinetic energy terms, which could lead to potential problems related to the presence of unphysical solution. Ghost instabilities are specific to higher order theories, like, for example, $f(R)$ gravity \cite{In6}.  However, as one can see from the first generalized Friedmann equation (\ref{F1}), in the present model the kinetic terms corresponding to the two fields are positive, and hence we do not expect the presence of ghost instabilities.
 
 The study of the linear stability and of the linear cosmological perturbations is another important field of research that gives important insights into the structure and viability of gravitational theories. For the case of nonminimally coupled scalar field models a detailed analysis of the stability and of the growth of linear perturbations was performed in \cite{In8}, by using a unified approach for the study of the stability and perturbations evolution. The stability and the growth of the linear perturbations for the exponential potential $V=V_0 e^{-\lambda \phi}$, and for the power law potential of the form  $V(\phi)\propto \phi ^n$. By assuming that the contribution of the Weyl boundary can be approximate by a barotropic fluid, the results of \cite{In8} can be also extended to the present model, to obtain at least some qualitative estimations of the stability properties. But the in depth investigation of the linear perturbations in the model requires a detailed mathematical and numerical analysis.       
 
 In particular, our results may also prove to be relevant for the warm inflationary model considered in \cite{warmweyl}, in which radiation creation  occurs during an accelerated phases of evolution of the Universe through the interplay and interaction of two distinct fields, a dissipative scalar field, and the boundary scalar field, respectively. While the Weyl scalar field is fixed unambiguously by the evolution equations of the adopted radiation creation model, the behavior of the scalar field is determined by its selfinteraction potential $V(\phi)$, which cannot be determined from the constraints on the warm inflationary model itself. Big Bang Nucleosynthesis offers a powerful method of obtaining restrictions of the considered forms of the scalar field potential through a direct comparison with observational data, via the consideration of the abundances of the primordial elements. Our analysis, based on a systematic statistical analysis, has provided some important information on the nature of the scalar field that may have triggered both the early accelerating expansion of the Universe, as well as the creation of matter. In particular, in the presence of the boundary terms, a simple quadratic potential $V(\phi)\propto \phi^2$ of a dissipative scalar field provides the best fit to the data. Exponential potentials also lead to good physical results. Hence, our investigations can provide the necessary statistical and theoretical tools that could lead to the possibility of the extensive testings of the various cosmological models proposed for the description of the dynamics and evolution of the early Universe. 

\begin{acknowledgments}
	We would like to thank the anonymous reviewer for comments and suggestions that helped us to significantly improve our manuscript. The work of T.M.\ was supported by a research fellowship funded by Babe\c{s}-Bolyai University of Cluj-Napoca.
\end{acknowledgments}

\end{document}